\documentclass{emulateapj}
\usepackage{aas_macros}
\usepackage{graphicx}
\usepackage{epstopdf}
\usepackage{float}

\newcommand{\be}{\begin{equation}}
\newcommand{\ee}{\end{equation}}
\newcommand{\ba}{\begin{eqnarray}}
\newcommand{\ea}{\end{eqnarray}}

\newcommand{\bnabla}{\mbox{\boldmath$\nabla$}}

\newcommand{\bOmega}{\mbox{\boldmath$\Omega$}}

\newcommand{\nn}{\mbox{} \nonumber \\ \mbox{} }

\begin{document} 

\title{Spin and Magnetism of White Dwarfs}
\author{Yevgeni Kissin}
\affil{Department of Astronomy and Astrophysics, University of Toronto, 50 St. George St., Toronto, ON M5S 3H4, Canada}
\author{Christopher Thompson}
\affil{Canadian Institute for Theoretical Astrophysics, 60 St. George St., Toronto, ON M5H 3H8, Canada}

\begin{abstract}
The magnetism and rotation of white dwarf (WD) stars are investigated in relation to a hydromagnetic dynamo
operating in the progenitor during shell burning phases.  The downward pumping of angular momentum
in the convective envelope, in combination with the absorption of a planet or tidal spin-up from a binary
companion, can trigger strong dynamo action near the core-envelope boundary.  Several arguments point to the outer core as the source
for a magnetic field in the WD remnant:  the outer third of a $\sim 0.55\,M_\odot$ WD is processed during the shell 
burning phase(s) of the progenitor; the escape of magnetic helicity through the envelope mediates the growth of (compensating) 
helicity in the core, as is needed to maintain a stable magnetic field in the remnant; and the intense 
radiation flux at the core boundary facilitates magnetic buoyancy within a relatively thick tachocline layer.
The helicity flux into the growing core is driven by a dynamical imbalance with a latitude-dependent 
rotational stress.  The magnetic field deposited in an isolated
massive WD is concentrated in an outer shell of mass $\lesssim 0.1\,M_\odot$ and can reach $\sim 10\,$MG.
A buried toroidal field experiences moderate ohmic decay above an age $\sim 0.3$ Gyr, which may lead to growth 
or decay of the external magnetic field.  
The final WD spin period is related to a critical spin rate below which magnetic activity shuts off,
and core and envelope decouple;  it generally sits in the range of hours to days.  WD periods ranging up to 
a year are possible if the envelope re-expands following a late thermal pulse.
\end{abstract}

\section{Introduction}\label{s:intro}

The magnetic field and spin of a white dwarf (WD) are inherited from its progenitor star.  Our interest here is 
in how magnetism and rotation evolve together within the progenitor, in response to an active 
hydromagnetic dynamo.  We focus on the post-main sequence phase of stellar evolution, when the material comprising the outer 
layers of the WD is deposited by a strong inflow from a hydrogen-rich convective envelope,
through burning shell(s), into the stellar core.  The presence or absence of a magnetic field at the 
surface of the WD remnant depends on the ability of the inner envelope to sustain a magnetic field.  

Rotationally-driven dynamo activity on the red giant (RGB) and asymptotic giant (AGB) branch has received 
only limited attention (e.g. \citealt{blackman01, nordhaus08}), due to a prevailing assumption that the radial angular velocity profile 
$\Omega(r)$ in the convective envelope will mirror that observed in the Sun.  Although localized magnetic field amplification in the outer, 
slowly rotating envelope may result in weak surface magnetic fields
\citep{dorch04,auriere10}, a dynamo that manifests itself through strong chromospheric and coronal emission
appears to depend on an interaction with a binary stellar companion, as in RSCvn stars (e.g. \citealt{moss91}).
Magnetic activity appears to be prevalent during the first stages of post-MS expansion, and during core helium
burning \citep{auriere14}, but the sampling of more expanded giants is very limited.  

On the other hand, the angular velocity profile in a giant is less sensitive to boundary conditions than it is in the 
Sun, given the extreme depth of the envelope.  
Here we are guided by i) recent measurements of rapid core rotation in subgiants and core He burning stars 
using asteroseismic measurements by Kepler \citep{becketal2012,mossetal2012}; 
ii) measurements of a strong decrease of surface rotation speeds in subgiants as they evolve toward the RGB \citep{SchrP1993}; 
and iii) anelastic simulations which show a tendency to nearly uniform specific angular momentum in a deep convective envelope with slow rotation \citep{brunp2009}.  

In a companion paper (\citealt{KissT2015}, hereafter paper I), 
we show that the Kepler measurements
are consistent with a tight coupling between the radiative core and convective envelope, combined with 
some inward pumping of angular momentum deep into the envelope.  The favored rotation profile has
$\Omega(r)$ transitioning between $\propto r^{-1}$ in the inner envelope to $\propto r^{-2}$ in the outer and 
more slowly rotating layers.  The depth of the transition point depends on the size of the star and its angular momentum.

An important consequence of such an inward-peaked rotation profile is that the threshold for a rotationally
driven dynamo may be reached in the inner envelope, but {\it not} near the stellar photosphere.
A key figure of merit is the Coriolis parameter  
\be\label{eq:coriolis}
{\rm Co} \equiv \Omega \tau_{\rm con} \equiv {\Omega\ell_P\over v_{\rm con}}
\ee
as measured at the base of the envelope; here $v_{\rm con}$ is the convective velocity and $\tau_{\rm con}$ is 
the timescale for convection over one pressure scale height.  Observations of MS stars indicate the
presence of dynamo activity when ${\rm Co} \gtrsim 1$ (corresponding to a Rossby number $\lesssim 2\pi$; 
e.g. \citealt{reinbb2009}).  The radial transition in the slope of the rotation profile in a deep convective 
envelope is postulated to occur where ${\rm Co}$ approaches unity, with the most rapidly rotating zone lying
below this transition.   

Inward pumping of angular momentum can also reduce
the need for magnetically driven spindown in subgiants (Paper I).

\subsection{Remnant Magnetism: Dependence on Stellar Mass, Binarity and Age}

Three hypotheses have dominated previous discussions of the origin of magnetic fields in WD stars:
the conservation of magnetic flux from the main sequence (MS) progenitor \citep{angebl1981,ferrw2005}; 
the amplification of such a seed field by convective episodes in the core of the progenitor star \citep{RudeS1973}; 
and the transient amplification of the field during a merger event \citep{ToutWLFP2008,NordWSMB2011}.  

We begin our investigation by reviewing some constraints on each of these hypotheses; and 
then summarize the arguments pointing to an amplification of the magnetic field 
during the post-MS expansion of the WD progenitor.  This process comprises 
three steps:  i) the advection of angular momentum inward through the convective envelope; ii)
the amplification of a magnetic field by a dynamo near the core-envelope boundary; followed by iii)
a flow of magnetic helicity into the growing core, which is sustained by a persistent angular velocity 
gradient and magnetic twist below the boundary.

Whether the inner envelope can support a dynamo turns out to depend on a combination of stellar properties. The core rotation rate is 
sensitive to the loss of angular momentum on the MS, and to any interaction with a planetary or stellar companion. Clearly solid rotation 
would imply ${\rm Co} \ll 1$ throughout the envelope of a star approaching the tip of the RGB or AGB; and therefore suggest the absence of a 
large-scale hydromagnetic dynamo.  
But when the envelope rotation profile is inwardly peaked, we find differing answers for isolated stars of 
initial mass $M_{\rm ZAMS}$ less or greater than $\sim 1.3\,M_\odot$, corresponding to the presence or absence of 
a strong magnetic wind on the MS.   Hence the greater importance of binary interaction for 
sustaining ${\rm Co} \gtrsim 1$ in solar-mass stars.  

Another consideration is the proportion of the material contained in the remnant WD that is deposited during single and double shell burning.  
This comprises about a third of the mass of the WD when $M_{\rm ZAMS} \sim 1$ ($M_{\rm wd}
\sim 0.55\,M_\odot$), but is strongly reduced by dredge-up (after core helium
ignitition and before the onset of thermal pulses) in the intermediate-mass
stars that are the progenitors of massive WDs.  In this second case, the magnetized mass shell turns out
to be thin enough that ohmic decay is initiated on a timescale $\sim 0.3$ Gyr (with some dependence
on the mass loss rate on the AGB). 

The source of a strong magnetic field ($1-10^3$ MG) in a WD therefore depends on both its age and mass.
Indeed, expanded sampling of WD magnetism calls into
question the flux-freezing hypothesis.
Whereas between $\sim 2$ and $\sim 10$ percent of isolated 
WDs are magnetic in this sense \citep{LiebBH2003,Liebetal2005} -- roughly similar to the proportion of chemically peculiar 
and magnetic A stars -- the proportion is higher in isolated massive WDs \citep{kepletal2013} and extends to 
$\sim 25$ percent in cataclysmic variables (CVs; \citealt{Gans2005}).  A large proportion of magnetic
WDs in short-period CVs also appear to have very strong magnetic fields, exceeding $10$ MG, which are much rarer
in the isolated WD population.

It has also been noted that magnetic WDs appear to be rare in binary
systems with a non-interacting MS companion \citep{Liebetal2005}, suggesting that binary interaction may be involved
in their genesis.  Previous theoretical ideas have centered around the impulsive growth of a magnetic field during a binary
merger \citep{NordBF2007,ToutWLFP2008,NordWSMB2011}.   Here our focus is on a much longer-lived process that
is sustained by the inward pumping of angular momentum.

A solar-mass star is slowly rotating at the end of the MS, so that the absorption of a giant planet significantly
spins up the star and may trigger dynamo activity \citep{livio02}.  The critical planetary mass that has this
effect near the base of the envelope depends on the angular velocity profile.  For 
a flat distribution of specific angular momentum at ${\rm Co} \lesssim 1$, we find that the absorption of even
a Neptune-mass planet can significantly spin up the core.   


A star more massive than $M_{\rm ZAMS} \sim 1.3\,M_\odot$ retains enough angular 
momentum at the end of the MS that only a planet more massive than Jupiter will significantly supplement its rotational angular momentum.  
Nonetheless we find that such an interaction can lead to stronger magnetic fields in the WD remnant.

Two properties of giants are especially relevant here.  The inward flow of matter across the core-envelope 
boundary allows a net transport of magnetic helicity, which has been shown to be a necessary ingredient in
the long-term stabilization of white dwarf magnetic fields \citep{brais2004}; and has also been implicated in the establishment of
solid rotation in the radiative core of the Sun \citep{GougM1998}.  We show that this helicity flux can be related to
the magnitude of the large-scale Maxwell torque that enforces (nearly) solid rotation in the outer core.

The outer core of an AGB star is also distinguished by an intense radiation flux. 
This greatly facilitates the buoyancy of a toroidal magnetic field
that is anchored in a tachocline layer below the convective envelope.
A thickness approaching $\sim 10$ percent of a scale height can maintain
contact with the envelope over a dynamo period $\sim 10-10^2 \tau_{\rm con}$.  This suggests that a given level
of magnetic activity (measured e.g. by the ratio of Maxwell to Reynolds stresses) may be achieved with a lower 
Coriolis parameter in an AGB envelope than in a solar-type MS star.  Magnetically induced mixing of core
material into the envelope \citep{nordhaus08} could also be significantly enhanced by this effect.

\subsection{Remnant Spin}

The measured spins of WDs are strongly inconsistent with solid rotation in the envelope of an AGB star, combined with 
a tight core-envelope coupling.  But even if we allow for inward convective pumping of angular momentum, the final 
spin rate of a WD remnant is sensitive to the heavy loss of mass during the transition to the 
post-AGB phase.\footnote{We thank Peter Goldreich for emphasizing this point to us.}  
   
A decoupling of the rotation of core and envelope is only feasible
when the large-scale poloidal magnetic field drops below the equivalent of $10^2-10^3$ G in the WD during
the brief, thermally pulsating AGB phase.  (Much weaker fields will couple the core and envelope of a subgiant
or core helium burning star; Paper I.)  This
appears, at first sight, to be inconsistent with strong magnetism in the WD.
In fact, a magnetic coupling between core and envelope is naturally self-limiting:  once ${\rm Co}$ drops 
below a critical value, the dynamo operating in the inner envelope shuts off.  
A reasonable first guess (${\rm Co}_{\rm crit} \sim 0.1-0.3$) yields WD rotation periods in the range observed
in most WDs.  

As a result, the remnant rotation is {\it inversely} correlated with the net angular momentum
in the star:  the slower the rotation of the AGB envelope, the sooner the decoupling between core and envelope,
when the core has a larger radius and moment of inertia.  The weakly magnetized
surface layer of the WD is very thin, and we show that the magnetic field below quickly diffuses to the photosphere. 

Our most surprising finding is that WDs with very long ($\sim$ yr) rotation
periods originate from stars which spin relatively rapidly
near the end of the AGB phase.  Then the rotation of core and envelope
remain magnetically coupled during the contraction of the envelope.  
The endpoint is sensitive to details of how the envelope is expelled:  in particular, to
whether the envelope experiences a re-expansion following a late thermal
pulse.  This taps most of the remaining angular momentum of the core.
Without such a re-expansion, the final rotation period can be shorter than a day.

It should be kept in mind that anisotropic contraction of the envelope can contribute to white dwarf rotation 
\citep{Spru1998}.  Indeed, it may be needed to generate centrifugally supported disks that provide seeds for
metal enrichment of white dwarf atmospheres.

\subsection{Stellar Models and Planetary Interaction}

In order to evaluate various pieces of the physics, we employ two stellar models of solar metallicity and initial mass
$M_{\rm ZAMS} = 1\,M_\odot$ and $5\,M_\odot$.  Both models are constructed using the MESA code \citep[version 5527]{paxtetal2011}, with 
rotational degrees of freedom turned off.  The normalization
of the mass-loss rate on the AGB, and the masses of the WD remants, were fixed by matching the latest initial-to-final
mass relation obtained from open clusters \citep{Dobbetal2009}.  The coefficient entering into the formula of
\cite{bloe1995} was taken to be $\eta_B = 0.05$.  The corresponding masses of the WD remnants are $0.55\,M_\odot$ and 
$0.87\,M_\odot$.  

The rotational evolution was followed in post-processing using the rotation models
described in Section \ref{s:rotation}.  The loss of angular momentum due to mass loss on the RGB and AGB was
taken into account, along with the interaction with a companion planet of a range of masses:  Earth, Neptune, Jupiter
in the $1\,M_\odot$ model, and even heavier companions in the $5\,M_\odot$ model.  In the first case, the companions were
started at a range of semi-major axes and orbital eccentricities.  The orbit of the planet,
and the exchange of angular momentum with the star, were computed taking into account tidal drag
and the gravitational quadrupole that is generated by convection in the giant envelope.
Further details are given in Paper I.

\subsection{Plan of the Paper}

Section \ref{s:basic} and \ref{s:WD_magnetism} discuss several processes that are relevant to the growth and maintenance of a stable magnetic 
field in the core of a post-MS star, and which determine the spin angular momentum trapped there at the end of the AGB.  
Section \ref{s:finalB} presents detailed estimates of WD dipole magnetic fields.
Section \ref{s:WD_rotation} analyzes the magnetic decoupling of core and envelope near the tip of the AGB, and provides
estimates of the final WD spin.  Ohmic transport through a thin, unmagnetized surface layer in a WD, and 
the thicker magnetized layer below it, is considered in Section \ref{s:B_field_emergence_decay}.  

Various outstanding issues -- such as the sensitivity of 
our results to the assumed angular velocity profile in the progenitor envelope, the impulsive growth of a magnetic field during a merger, 
and effects associated with the anisotropic contraction of the residual AGB envelope -- are addressed in the 
concluding Section \ref{s:conclusions}.  The Appendix gives further details of the accumulation of magnetic helicity in the stellar core.

\section{Rotation and Magnetism in a Rapidly Evolving Giant Star}\label{s:basic}

Here we summarize the rotation model for giant stars that was developed in Paper I.  We then 
review some physical constraints on generating sustainable white dwarf magnetic fields, with a focus on 
the accumulation of magnetic helicity during the rapid growth of the hydrogen-depleted
core of a giant star. We also summarize
the relative effectiveness of different convective episodes at sourcing WD magnetic fields.
In this regard, there are some significant structural differences between the evolving cores of solar-mass and of intermediate-mass stars.  

\subsection{Combined Effect of Convection and Poloidal Magnetic Field on Angular Momentum Redistribution}\label{s:rotation}

Our understanding of the origin of WD rotation has been impeded by an incomplete knowledge of mixing processes 
in the radiative parts of the progenitor.  Even a very weak poloidal magnetic field (corresponding to a magnetic
field $\ll$ MG in the WD under the assumption of magnetic flux freezing) will enforce solid rotation in the evolving
core.  

A popular approach recently has been to assume that such a large-scale poloidal field is initially
absent, and to imagine that the radial magnetic field that is needed for angular momentum transport is generated by
a current-driven (`Tayler') instability of a predominantly toroidal field (\citealt{Spru2002,CantMBCP2014}, and references therein).
This leads to an incomplete coupling between core and envelope.  We argue that even an inefficient dynamo
process operating near the core-envelope boundary will easily short-circuit this effect.  

We develop, as a working alternative, a simple and deterministic model of angular momentum redistribution 
in evolving stars.  This model applies to stars which retain (or gain) enough angular momentum to support
some dynamo activity near the core-envelope boundary.  

\vskip .1in \noindent
1. The outer core is assumed to rotate as a solid body,
as enforced by large-scale Maxwell stresses, and to co-rotate on the average with the inner envelope.  

\vskip .1in \noindent
2. Inhomogeneous rotation is sustained mainly in convective regions of the star, especially the deep outer
envelope that forms following the completion of core H and He burning.  
\vskip .1in

The rotation profile that is formulated and calibrated in paper I can be summarized as follows.  
Where the Coriolis force can 
be neglected, outside some transition radius $R_c$, the envelope maintains uniform specific angular momentum,
\be\label{eq:omouter}
\bar\Omega(r) = \bar\Omega(R_c) \left({r\over R_c}\right)^{-2};  \quad r > R_c.
\ee
Here $\bar\Omega$ is the angular velocity averaged over a spherical shell.  We identify 
$R_c$ with the base of the envelope,
\be
R_c = R_{\rm benv}  \quad (\bar{\Omega} \tau_{\rm con} < 1\;{\rm at} \; R_{\rm benv}),
\ee
when the angular momentum of the star is small enough that $\bar\Omega(r) \tau_{\rm con} < 1$ throughout the envelope.
Throughout this paper, the subscript `benv' refers to the base of the envelope.

When the star has more angular momentum, we maintain the profile (\ref{eq:omouter}) in the
outer part of the convection zone, and consider
\ba\label{eq:ominner}
\bar\Omega(r) &=& \bar\Omega(R_c) \left({r\over R_c}\right)^{-\alpha} \nn
&=& {{\rm Co}_{\rm trans}\over \tau_{\rm con}(R_c)} \left({r\over R_c}\right)^{-\alpha};  \quad
R_{\rm benv} < r < R_c
\ea
in the inner envelope. The Coriolis parameter at the transition radius ${\rm Co}_{\rm trans}$ is taken to
be unity, and the radial index $1 < \alpha < 3/2$. As the angular momentum increases $R_c$ moves outward, 
and may reach the surface of the star, in which case ${\rm Co}$ is larger than unity everywhere in the envelope.

As we discuss in Paper I, $\alpha=1$ best matches the observed core rotation periods of sub-giant and helium burning stars, 
which we use to calibrate our rotational model. However, in stars of much greater size, such as those approaching the tips of
the RGB and AGB, the core is more centrally concentrated and the gravitational field in the inner
envelope, $g(r) \propto r^{-\beta}$, steepens from $\beta = 1$ to $\beta = 2$.  Then the value of $\alpha$ that results from the inward transport of
angular momentum by deep convective plumes increases from $(1+\beta)/2 = 1$ to $3/2$.   It is during this more advanced phase of
evolution that the helicity flux into the core is expected to peak, and the rotational fate of the WD is determined. 
We therefore use $\alpha=3/2$ in our analysis.

\subsection{Magnetic Helicity Growth in Giant Cores}\label{s:htransport}

Net magnetic helicity ${\cal H} = \int{\bf A}\cdot{\bf B}\,dV$ appears to be an essential ingredient for maintaining long-term
magnetostatic stability in a static and entirely fluid star \citep{brais2004}.  Here ${\bf B}$ is the magnetic field
and ${\bf A}$ the vector potential from which it is derived. ${\cal H}$ is a topological charge in a perfectly conducting
fluid without boundary.  For example, when a closed poloidal magnetic loop carrying flux $\Phi_p$ is also twisted through an angle
$\psi$, the helicity is ${\cal H} = \Phi_p\Phi_\phi = \psi\Phi_p^2$, where $\Phi_\phi$ is the toroidal flux.  ${\cal H}$ evolves
only on a very long ohmic timescale in an extended astrophysical fluid -- unless there is an exchange of magnetic twist with 
the exterior of the fluid, or with a medium of low electrical conductivity.  

The accumulation of helicity considered here is associated with rapid changes in the structure of the star.
We expect this to be possible in the outer core and inner envelope of a giant star, 
since this zone is connected to the surface of the star by rapid convective motions.  Core convection has a topology 
less favorable to the accumulation of ${\cal H}$, as does the erasure of differential rotation by the
stretching of magnetic field lines in the interior of a radiative star.  A simplified time-dependent model of 
an internal shear-driven dynamo has been developed by \cite{WickTF2014}, but does not address the growth of ${\cal H}$.

Such a secular accumulation of magnetic helicity -- although occuring relatively rapidly near the tips of
the RGB and AGB -- is nonetheless a gradual process when measured over an individual dynamo cycle.  There has also
been longstanding interest in the role that helicity conservation may play in limiting the exponential growth of
a seed magnetic field in a turbulent magnetofluid \citep{gruzinov94,bf00a,vishniac01}.  In mean-field dynamo models, this involves 
a suppression of the $\alpha$ effect, e.g. the toroidal electromotive force which sources a large-scale poloidal magnetic field.

This limitation on fast dynamo action could be removed by the transport of
magnetic helicity on a relatively short timescale, comparable to the rotation period. Such transport may involve
the removal of twist across the surface of the star \citep{bf00b}, or alternatively an internal flow of 
helicity between rotational hemispheres \citep{vishniac14}.  In either of these
cases, rapid dynamo growth is not directly dependent on changes in the net stellar helicity, obtained by summing the contribution from both 
hemispheres. The longer, secular effect being considered here necessarily
depends on a breakdown of reflection symmetry between magnetic hemispheres; and, because it involves a change in an integral property of the 
star, must depend on the transport of twist across the stellar surface.

The connection between helicity growth and mass inflow can be examined with a simple model of a toroidal
magnetic twist superposed on a radial magnetic field, all subjected to a radial MHD flow of speed $v_r$. 
(The sign of this flow, relative to the convective boundary, fluctuates during thermal pulses on the AGB 
but is negative on the average.)  

Within a sphere coinciding with $R_{\rm benv}$, one finds 
\ba\label{eq:dhdt}
{d{\cal H}\over dt} &=& {d\over dt}\int {\bf A}\cdot{\bf B}\, dV \nn
&=& 
{\partial{\cal H}\over \partial t} + {dR_{\rm benv} \over dt}\int dS A_\phi B_\phi
\;=\; -\delta v_r\int dS\,A_\phi B_\phi\nn
\ea
(Appendix \ref{s:twist}). The gauge
\be
A_\phi(\theta,r) = {1\over r\sin\theta}\Phi_r(\theta)
\ee
is convenient, where 
\be \label{eq:flux}
\Phi_r(\theta) = \int_0^\theta \sin\theta' r^2B_r(\theta') d\theta'
\ee
is $(2\pi)^{-1}$ times the radial magnetic flux between the rotation axis and polar angle $\theta$.
Clearly $\Phi_r(\pi) = 0$.

One notices that the surface integral appearing in Equation (\ref{eq:dhdt}) has the dimensions of a torque:
it is comparable in magnitude to the Maxwell stress $B_r B_\phi/4\pi$, multiplied by the lever arm
$r\sin\theta$, and integrated over the core-envelope interface.  Even if the net torque acting on 
the core boundary vanishes, $d{\cal H}/dt$ may remain finite as a consequence of the different
angular dependences of $A_\phi$ and $B_r r\sin\theta$.

We are interested here in the case where $B_r B_\phi/4\pi$ is the dominant stress driving the core
material toward solid rotation.  Then we can estimate the helicity flux into the core once we know
the source of the Maxwell torque.  The largest effect turns out to be the latitude dependence of
the Reynolds stress that is applied to the core at the base of the convective envelope.  

We may estimate this stress by considering a dynamo process at work in a tachocline layer 
right below the convective boundary.  
Convection above the tachocline supports strong latitudinal variations in $\Omega$ which impose a radial mis-match 
$\Delta\Omega$ with the mean core rotation $\Omega_{\rm benv} = \bar\Omega(R_{\rm benv})$.  
The tachocline can therefore be identified with the radial layer (of thickness $\Delta r$)
in which $\Delta\Omega$ is concentrated, 
\be
\bnabla\Omega \simeq {\partial\Omega\over \partial r}\hat r
                 \sim {\Delta\Omega\over \Delta r} \hat r.
\ee
Our focus here is on the parts of the tachocline where $\partial\Omega/\partial r > 0$
(in the Sun, it is the equatorial band at latitudes $\lesssim 45^\circ$).  
The magnetorotational instability \citep{BalbH1994} is excited where $\partial\Omega/\partial r < 0$,
and is sustained by rapid radiative diffusion even in the presence of stable radiative convective
equilibrium \citep{menou04}.

The poloidal magnetic field threading this part of the tachocline is wound up linearly on
timescales short compared with the dynamo period $P_{\rm dyn}$, which we normalize as
$N_{\rm dyn}\Delta\Omega^{-1}$. We limit the magnetic field strength in the tachocline 
by noting that a radial magnetic field stronger than 
\be
B_r \sim (4\pi \rho)^{1/2} {\Delta r\over P_{\rm dyn}}
\ee
will act to redistribute angular momentum in a layer of density $\rho$ and thickness $\Delta r$, thereby
shorting out the angular velocity gradient.  
The ratio of toroidal to poloidal field components  is obtained from the linear winding term in the
induction equation,
\be\label{eq:bphir}
{2\pi\over P_{\rm dyn}} B_\phi \sim {r \Delta\Omega \over \Delta r} B_r.
\ee
One finds that the toroidal magnetic field is in approximate equipartition with 
the rotational motions at the base of the convective envelope,
\be\label{eq:bphieq}
B_\phi^2 \sim 4\pi \rho \left(r\frac{\Delta\Omega}{2\pi}\right)^2 \sim 
4\pi \left({{\rm Co}\over 2\pi}{r\over\ell_P}\frac{\Delta\Omega}{\bar{\Omega}}\right)^2\rho v_{\rm con}^2 .
\ee
The Maxwell stress works out to 
\be\label{eq:reyn}
{B_r B_\phi\over 4\pi} \sim {\rho r \Delta r (\Delta\Omega)^2 \over 2\pi N_{\rm dyn}}.
\ee
We emphasize that this estimate of the stress has been obtained from dynamical considerations,
and side-steps the challenging question of how the $\alpha$ effect is actually manifested in a
dynamo operating near a radiative-convective boundary.  For recent approaches to this problem,
in the context of lower-luminosity MS stars, see \cite{bt15} and references therein.

We note that $\Delta\Omega \sim \bar\Omega$ is seen in anelastic calculations of deep and rapidly rotating
(${\rm Co} > 1$) convective envelopes \citep{brunp2009}, and is suggested by the analytic treatment of
angular momentum pumping in Paper I.
The latitudinal gradient is milder in the Sun, where the equator rotates more rapidly than the poles, 
$(\Omega_{\rm eq}-\Omega_{\rm pole})/\bar{\Omega} \sim 0.1$; but the Solar convective envelope also has a very
different aspect ratio.

Now consider the helicity flux integrated over one rotational hemisphere.  Even in a state of rotational 
equilibrium -- where the angle-integrated Maxwell torque vanishes -- the integral (\ref{eq:dhdt}) generally
remains finite.   Notice also that it is the product
of $B_r$ and $B_\phi$ that is regulated according to Equation (\ref{eq:reyn}).  The Maxwell
stress does not depend on the sign of $B_r$ that is imposed by a dynamo operating near the core-envelope
boundary, nor does it depend directly on the coherence of the magnetic field.

The sum of the two hemispheric integrals for $d{\cal H}/dt$ would vanish if the magnetic field were 
reflection-symmetric about the rotational equator.  But in general it is not:  the distribution of 
radial magnetic flux across the northern hemisphere need not precisely mirror that threading the 
southern hemisphere.  (Only the net spherical flux vanishes.)   Such an incomplete hemispheric
cancellation allows a net flow of magnetic twist into the core. 

Other sources of helicity flow can be considered.  The inflow through the core-envelope boundary
itself drives $\partial\Omega/\partial r$, but on a much longer timescale than the rotation period.

The WD magnetic fields that result from these two sources of 
${\cal H}$ are compared in Section \ref{s:WD_magnetism}.

\subsection{Relative Importance of \\ Different Convective Episodes}

The magnetic field of a WD is influenced by a series of convective episodes in the progenitor.  One measure
of the relative importance of different convective episodes is provided by the convective Mach number
${\cal M}_{\rm con} = v_{\rm con}/c_s$ (here $c_s$ is the adiabatic speed of sound), which in a stellar core can be
estimated as 
\be
{\cal M}_{\rm con} \sim \left({R_{\rm core} v_{\rm con}^2 \over GM_{\rm core}}\right)^{1/2}.
\ee
As the core continues to contract during later stages of nuclear burning, its magnetic energy and gravitational 
energy scale in the same way with central density. To the extent that the generated Maxwell stress is limited by the convective stress, e.g. 
$B_\phi \lesssim (4\pi \rho)^{1/2} v_{\rm con}$, episodes of greater 
${\cal M}_{\rm con}$ are capable of generating stronger magnetic fields.  

One encounters much larger ${\cal M}_{\rm con}$ on the giant branches ($\sim 10^{-3}-10^{-2}$) than during steady core H or He
burning in intermediate-mass stars ($\sim 10^{-5}-10^{-4}$).  
It is also worth noting that the intense period of convection during the AGB phase is also the shortest in duration.  That implies
a weaker cancellation in the magnetic helicity that accumulates in the core, due to a shifting distribution of poloidal magnetic flux
(see Section \ref{s:htransport}).

Magnetic fields that are generated during a certain convective phase will remain trapped unless exposed to later 
stages of convection, or the exterior mass shells are expelled.  As we examine in Section  \ref{s:B_field_emergence_decay},
ohmic diffusion in the WD remnants of intermediate-mass stars is restricted to a relatively thin magnetized layer.
The relevance of dynamo activity on the RGB, horizontal branch and AGB is determined, in good part,
by the maximum inward penetration of the convective envelope:  one requires $M(R_{\rm benv}) \leq M_{\rm wd}$.
Only in relatively massive progenitors ($M_{\rm ZAMS} \gtrsim 4\,M_\odot$) does MS core convection
extend to the mass boundary of the WD remnant, where it can influence the visible magnetic field; we ignore this effect here.  

The core helium flash occuring in stars of mass $< 2.3\,M_\odot$ is confined to a central subset of the eventual WD material.
Because a star does not become fully convective during the flash (e.g. \citealt{bildsten12}), 
the magnetic helicity of the convecting core material is essentially conserved.  A similar conclusion applies to the brief
convective layers that are triggered by thermal pulses in the helium-burning shell near the tip of the AGB.

\subsubsection{Stars with Minimal Dredge-up of Helium ($M_{\rm ZAMS} \lesssim 2.3\,M_\odot$)}

\begin{figure}[!]
\figurenum{1}
\epsscale{1.0}
\plotone{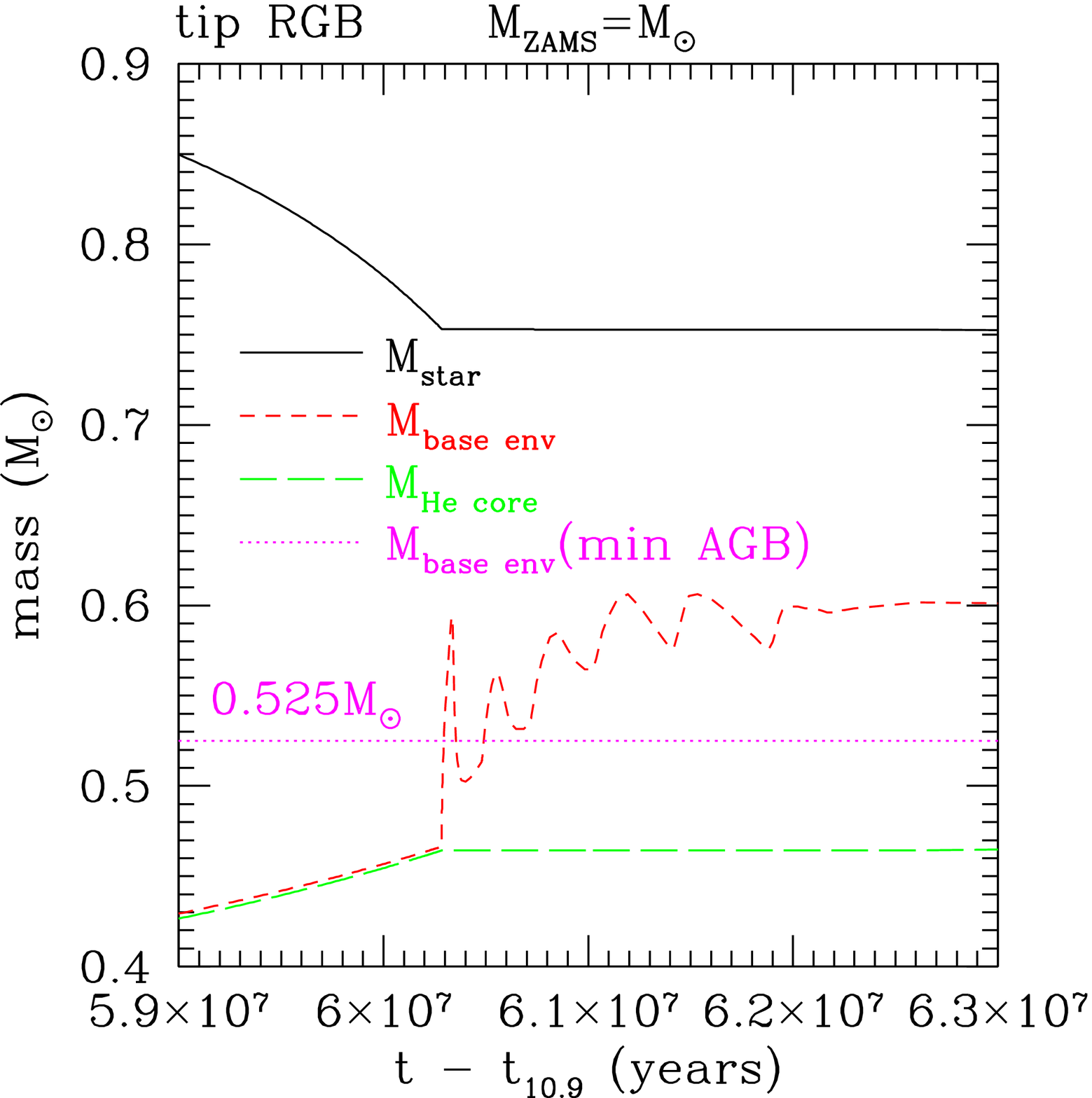}
\vskip 0.2in
\plotone{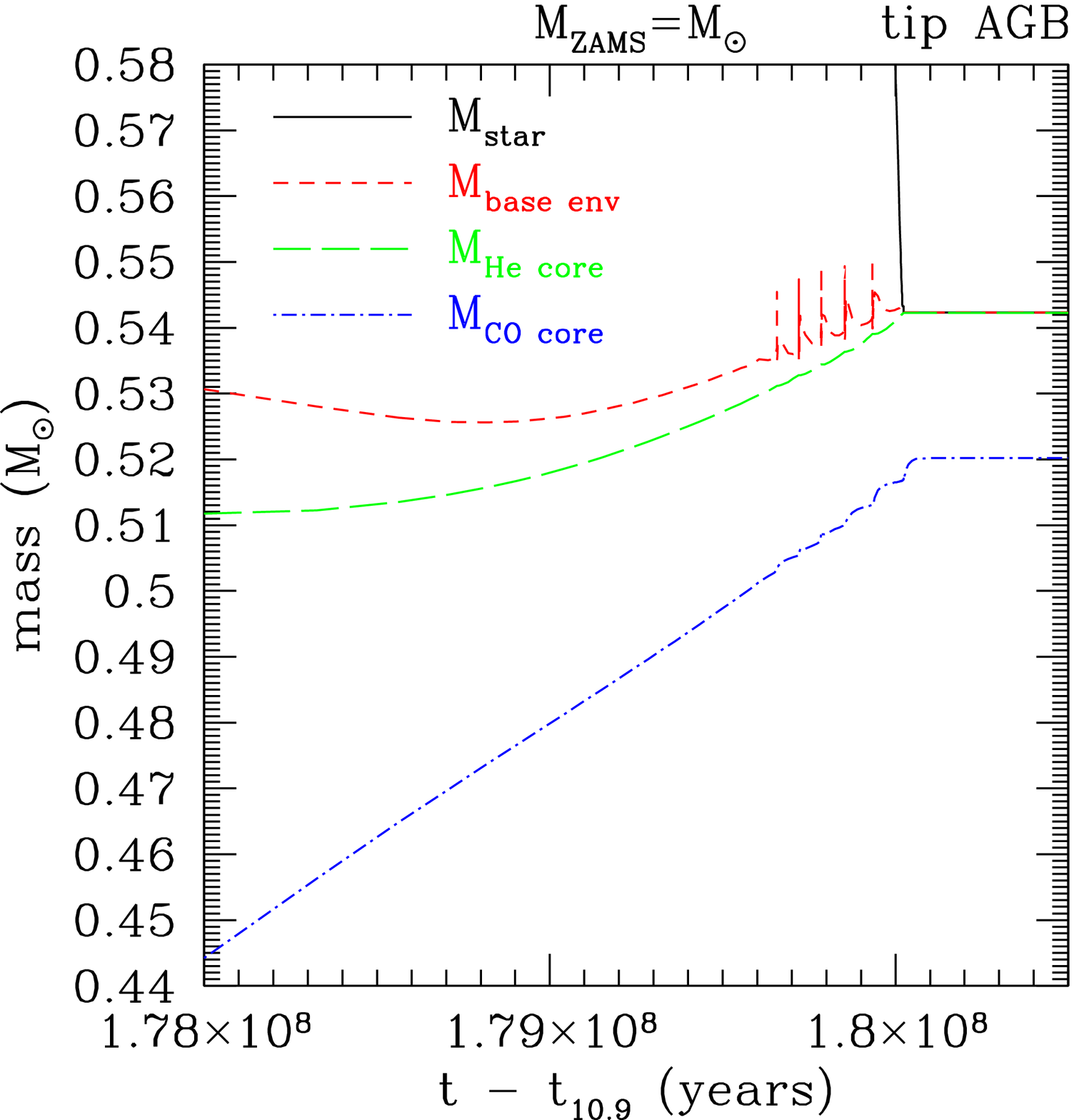}
\caption{Interior profile of $M_{\rm ZAMS} = 1 M_\odot$ star around the tip of the RGB and first stages of core
He burning (top panel) and during AGB phase (bottom panel). 
Dotted magenta line: maximum penetration of the convective envelope on the AGB ($0.525M_{\odot}$).   
A dynamo operating on the RGB deposits magnetized material within mass shells below this line; above it, the magnetic
field is sourced on the AGB.  Reference time $t_{10.9} \sim 12.37$ Gyr subtracts MS and early giant branch evolution
($R_{\star}(t_{10.9}) \sim 10.9R_{\odot}$).}
\vskip .2in
\label{fig:M_star_M_base_M_He_core_M_CO_core_TRGB_solar}
\end{figure}

In stars of initial mass $M_{\rm ZAMS} \sim 1\,M_\odot$, magnetic fields generated on the RGB are buried inside the WD remnant
under a layer of CO-rich material that is produced during double shell burning (Figure \ref{fig:M_star_M_base_M_He_core_M_CO_core_TRGB_solar}).
They are ohmically coupled on a timescale $\sim 10^8-10^9$ yr to a thinner magnetized shell that appears on the 
AGB (Section \ref{s:B_field_emergence_decay}).  

The ohmic timescale across the entire magnetized shell is longer than $\sim 10^9$ yr, meaning that any magnetic field
deposited near the center of the star during the pre-MS phase would influence the surface of the star only on a
timescale of a few Gyr.  

We conclude that the existence of young magnetized WDs with masses $\lesssim 0.6\,M_\odot$ points to dynamo activity in the convective envelope on the RGB and AGB.

\begin{figure}[!]
\figurenum{2}
\epsscale{1.0}
\plotone{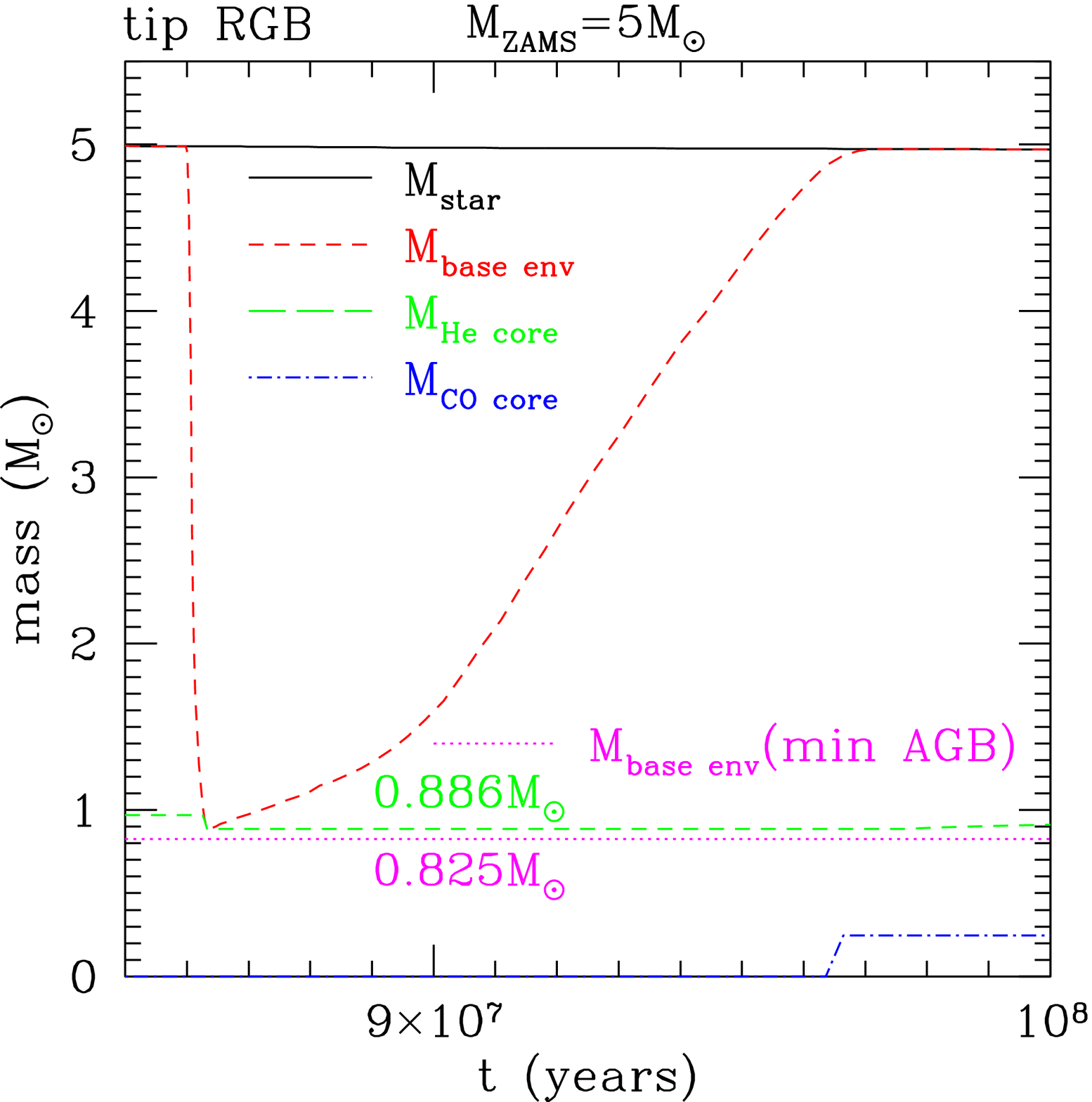}
\vskip 0.2in
\plotone{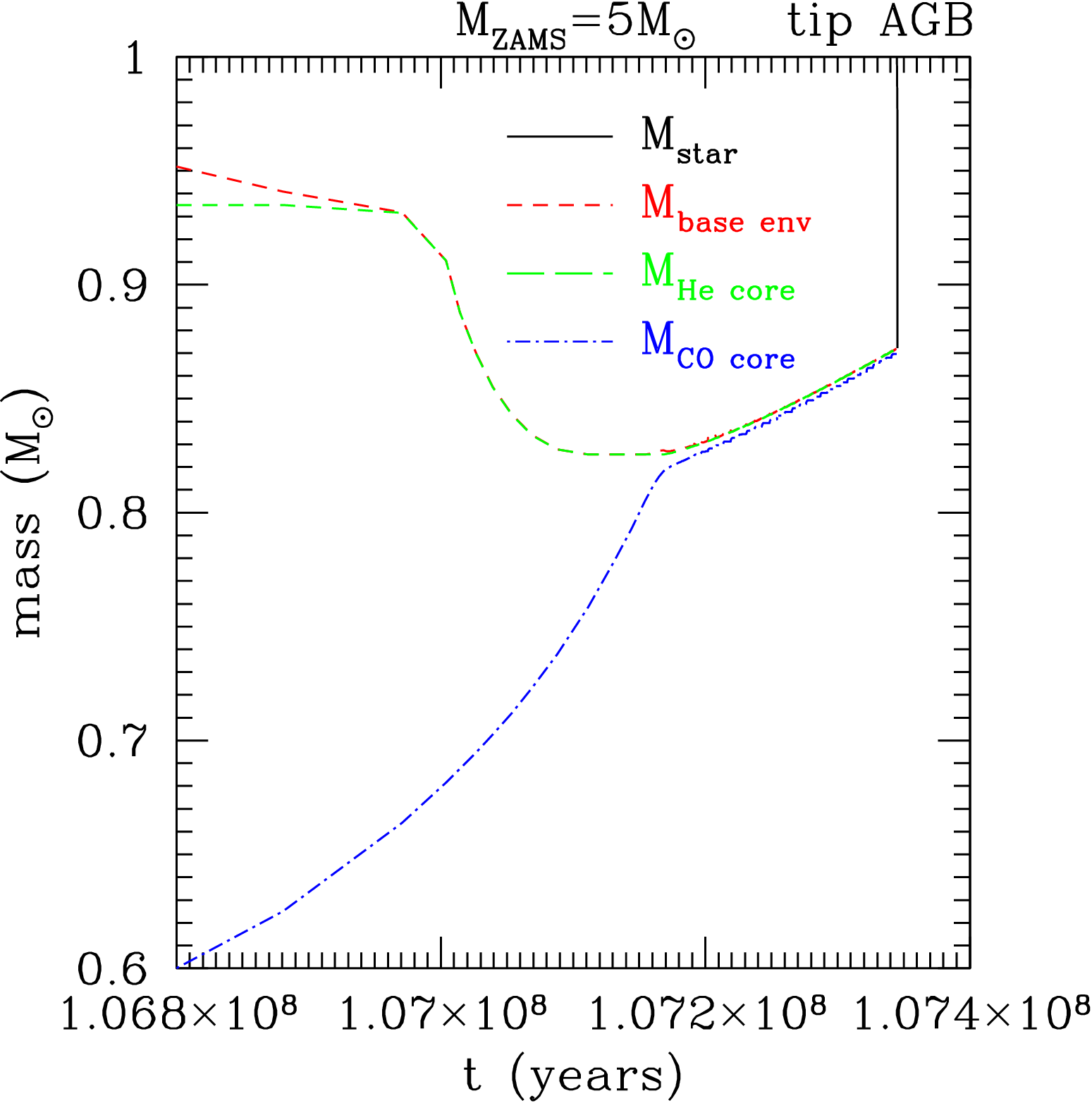}
\caption{Interior profile of a $M_{\rm ZAMS} = 5M_{\odot}$ star during RGB expansion and core He burning phase (top panel) and AGB phase 
(bottom panel).  The deepest penetration of the convective envelope on the RGB remains outside the mass boundary of the WD remnant 
($\sim 0.87M_{\odot}$).  Dredge-up of helium occurs on the early AGB;  the dotted magenta
line shows the corresponding maximum penetration of the convection.}
\vskip .2in
\label{fig:M_star_M_base_M_He_core_M_CO_core_TRGB_5M_o}
\end{figure}

\subsubsection{Intermediate-mass Stars ($M_{\rm ZAMS} \gtrsim 2.3\,M_\odot$)}\label{s:inter}

Stars of intermediate mass are distinguished from those of solar mass in that a dynamo operating
on the RGB cannot contribute directly to the WD magnetic field:  the outer parts of the helium core 
are dispersed by dredge up before the C/O core is assembled.  In addition, only a limited mass is
converted to carbon and oxygen by double shell burning (Figure \ref{fig:M_star_M_base_M_He_core_M_CO_core_TRGB_5M_o}).
The fraction of the WD mass contained by this outer shell is sensitive to the rate of mass loss on the 
thermally pulsating AGB.

The ohmic conduction time across this outer magnetized shell is several $10^8$ years. 
The surface magnetic field of the remnant of an isolated, intermediate-mass star is therefore expected to grow and
then gradually decay on this timescale (Section \ref{s:B_field_emergence_decay}).

\section{Magnetic Field Amplification Near Core-Envelope Boundary} \label{s:WD_magnetism}

We now investigate the growth of a magnetic field above and below the core-envelope boundary of a post-MS star.
There are two profound differences here with a solar-type star.  First, the radiative flux that is generated
by the burning shell(s) much exceeds the flux emerging from the core on the MS.  Second, hydrogen-rich material 
drifts downward rapidly through the convective boundary, as the core grows in mass.  This leads to a secular accumulation of magnetic
helicity in the core, as this inward drift is combined with a persistent magnetic twist.  Such a twist is generated when the magnetic field
couples the rotation of core and envelope.  In these respects, our considerations extend beyond existing dynamo theory.

\subsection{Critical {\rm Co} for Dynamo Action: \\ Effect of an Intense Radiation Flux in the Tachocline}\label{s:dynthresh}

Late-type MS stars generally are compact enough to sustain ${\rm Co} \gtrsim 1$ in their convective envelopes, as
are subgiants massive enough to have retained most of their natal angular momentum at the completion of core H burning.
Dynamo activity, as measured by chromospheric lines and coronal X-ray emission,
begins to decline when the Rossby number ${\rm Ro} \equiv P_{\rm rot}/\tau_{\rm con} = 2\pi/{\rm Co}$ is larger than about 0.1, 
corresponding to ${\rm Co} \lesssim 60$. Specifically in M-dwarfs, the closest MS analogs of giants, \cite{reinbb2009} observe
that surface magnetic flux scales as ${\rm Co}^2$ for ${\rm Co} \lesssim 60$.
There does not appear to be a sharp cutoff of magnetic activity with Coriolis parameter, as the simplest mean-field dynamo models would 
suggest.

The evidence therefore points to the presence of weak, but finite, dynamo activity at ${\rm Co} \sim 1$ in MS stars.   
It also begs the question as to whether the relative strength of the dynamo-generated magnetic field 
(measured in terms of the convective stress at a given Coriolis parameter) differs in giants and MS stars.

First it should be emphasized that a large-scale Maxwell torque that couples the rotation of the outer core to the 
envelope depends on a tiny poloidal field that is implanted from the dynamo layer.  We quantify this minimal
field in Section \ref{s:bseed}.

Some additional insight is provided by mean-field dynamo theory, which suggests that the dynamo number 
\be\label{eq:dynum}
D \equiv {\alpha \Omega r^3\over \nu_t^2} \sim \left({\alpha\over v_{\rm con}}\right) \left({r\over\ell_P}\right)^3
{\rm Co}
\ee
must exceed a threshold $D \gtrsim (2\pi)^3 \sim 10^2$ for magnetic field growth to be sustained \citep{Char2014}.  Here $\alpha$ is the 
parameter relating the toroidal magnetic field to the toroidal electromotive force, and $\nu_t$ is the turbulent
diffusivity.  The prospect for a dynamo is enhanced if the magnetic field can be anchored in a part of the star with
relatively low (but finite) $\nu_t$.  

This effect provides part of the motivation for considering a thin, stably-stratified layer below the convective
envelope as the locus of strong differential rotation, and a reservoir for a toroidal magnetic field
\citep{park1993}.  However, the sharpness of the onset of stable stratification below the envelope boundary has presented 
a difficulty for this hypothesis.  The solar convective motions have a relatively low Mach number 
(${\cal M}_{\rm con} \sim 10^{-4}$), and so the overshoot layer below the envelope is very thin.  A toroidal magnetic 
field reaching equipartition with the convective stresses overcomes the stable stratification in a layer that is 
not much thicker.

In Section \ref{s:buoyrad}, we show that the intense flux of radiation emanating from the core of a RGB or AGB star
will induce an upward drift of magnetized material, one that more than counterbalances the downward
flow to the burning shell(s).  A tiny seed field that is generated by the convective motions will sustain this feedback 
(Section \ref{s:bseed}).  The interchange between toroidal and poloidal fields will persist in a layer of considerable thickness
 $\sim (0.01-0.1)\ell_P$ over a dynamo cycle $\sim (10-10^2)\tau_{\rm con}$.  Convectively driven turbulence is strongly suppressed within 
this layer.



\begin{figure}[!]
\figurenum{3}
\figurenum{3}
\epsscale{1.0}
\plotone{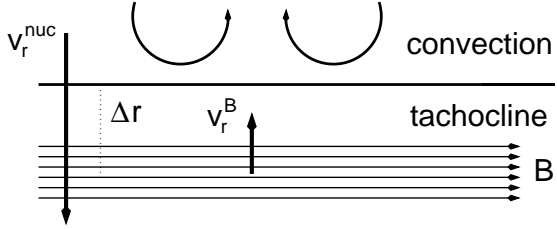}
\caption{A magnetic flux tube positioned in the tachocline layer experiences an upward drift that is driven by the intense radiation
flux through the tube.  This drift can be strong enough to counterbalance the downward flow of hydrogen-rich material
into the burning shell(s) below.}
\vskip .2in
\label{fig:fluxtube}
\end{figure}

\subsection{Magnetic Buoyancy in the Tachocline}\label{s:buoyrad}

\begin{figure}[!]
\figurenum{4}
\epsscale{1.0}
\plotone{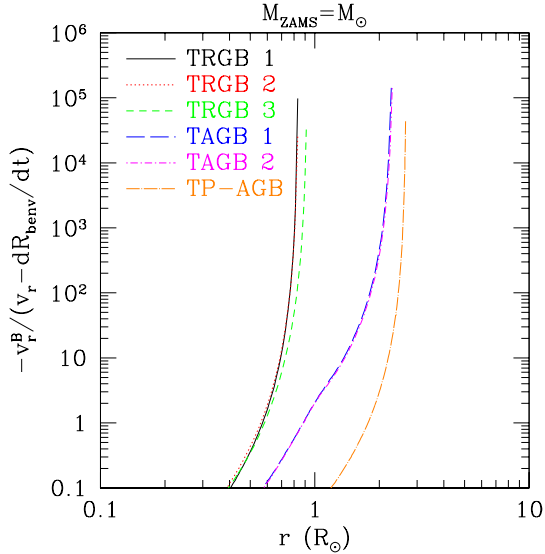}
\caption{Horizontal (toroidal) magnetic field anchored in a tachocline experiences radiatively driven buoyancy.
Upward drift speed $v_r^B$, Equation (\ref{eq:buoy}), is compared with the downward flow of material from the 
convective envelope into the growing radiative core.  Curves correspond to 
various timesteps near the tips of the RGB and AGB in our $1\,M_\odot$ model, and one taken from the thermally pulsating AGB.  The 
magnetic field is normalized to $B_\phi \sim (4\pi \rho v_{\rm con}^2)^{1/2}$, with $v_r^B$ proportional to $B_\phi^2$.
Buoyant drift accelerates toward the core-envelope boundary, due to the decrease in the Brunt-V\"ais\"al\"a frequency.
Here downflow is evaluated directly from the stellar model.}
\vskip .2in
\label{fig:vbuoy}
\end{figure}

We now examine the competition between the downward drift of hydrogen-rich material across the tachocline, and the upward
drift of magnetized material (Figure \ref{fig:fluxtube}).  Consider a horizontal (toroidal) magnetic flux tube, of a radial 
thickness comparable to the distance $\Delta r = R_{\rm benv}-r$ to the base of the convective envelope.   The flux tube
starts in buoyancy equilibrium, with a density equal to that of its unmagnetized surroundings.  The temperature deficit
in the flux tube (labelled `$B$') is
\be
{T_B - T\over T} = -{B_\phi^2\over 8\pi P} \ ,
\ee
which maintains a differential energy flux 
\be
\delta F_{\rm rad} \sim -{4ac\over 3\kappa\rho\Delta r} (T_B-T)T^3
\ee
into the flux tube.  This drives a buoyant rise of the flux tube at a speed $v_r^B$, which is determined by 
\be
\rho k_{\rm B}T v_r^B {dS\over dr} = -{dF_{\rm rad}\over dr} \sim {\delta F_{\rm rad}\over \Delta r}.
\ee
The background entropy gradient is usefully expressed in terms of the Brunt-V\"ais\"al\"a frequency
\be
N = \sqrt{g {\gamma-1\over \gamma} {dS\over dr}}, 
\ee
so that
\be\label{eq:buoy}
v_r^B \sim \left({B_\phi^2\over 8\pi P}\right) {F_{\rm rad}\over (\Delta r)^2 \rho N^2}.
\ee

The background flow of mass into the radiative core is 
\be
{dM_{\rm core}\over dt} = 4\pi R_{\rm benv}^2 \rho(R_{\rm benv})\left({dR_{\rm core}\over dt} - v_r^{\rm nuc}\right) \sim 
{L\over\varepsilon_i}.
\ee
After the star passes onto the thermally pulsating AGB, the relation between core growth rate and luminosity $L$ only holds after averaging
over pulsations.  The nuclear energy released per unit mass of new core material is $\varepsilon_1 = 6.3 X$ MeV/$m_u$ 
during the first step of converting material with hydrogen mass fraction $X$ to helium; and $\varepsilon_2 = \varepsilon_1 + 0.8$ MeV/$m_u$ 
during double shell burning.  The corresponding velocity is
\be
v_r^{\rm nuc} \sim -{F_{\rm rad}\over\rho \varepsilon_i}.
\ee

The ratio $v_r^B/|v_r^{\rm nuc}|$ is plotted in Figure \ref{fig:vbuoy} at various intervals during the RGB and AGB
evolution of our $1\,M_\odot$ model.  
One observes that the most strongly magnetized parts of the tachocline will
not be subducted if the magnetic pressure approaches the kinetic pressure at the base of the convective envelope.
It should be kept in mind that this effect depends on a difference in magnetization:  subduction of the less weakly 
magnetized parts of the tachocline must continue even where $v_r^B > |v_r^{\rm nuc}|$.  

Magnetic buoyancy has the effect of expanding the active dynamo region below $R_{\rm benv}$, by allowing a layer
with strong $d\Omega/dr$ -- the tachocline -- to remain in contact with the convection zone.\footnote{We do not address here whether the 
re-conversion of wound toroidal magnetic field back to the poloidal direction is concentrated in the tachocline, or instead 
the convection zone, as in Parker's 1993 model.}
We can estimate the size of this region by comparing the buoyant rise time $\Delta r/v_r^B(r)$ with the dynamo timescale $P_{\rm dyn}$ 
(set equal to $100\tau_{\rm con}$)

\begin{equation} \label{eq:Delta_r_buoyancy}
\Delta r \equiv R_{\rm benv}-r = v_r^B(r) \cdot P_{\rm dyn}.
\end{equation}
To invert this expression, we fit $v_r^B(\Delta r)$ to a power law in $\Delta r$, measuring it in two different radial zones.

\begin{figure}[!]
\figurenum{5}
\plotone{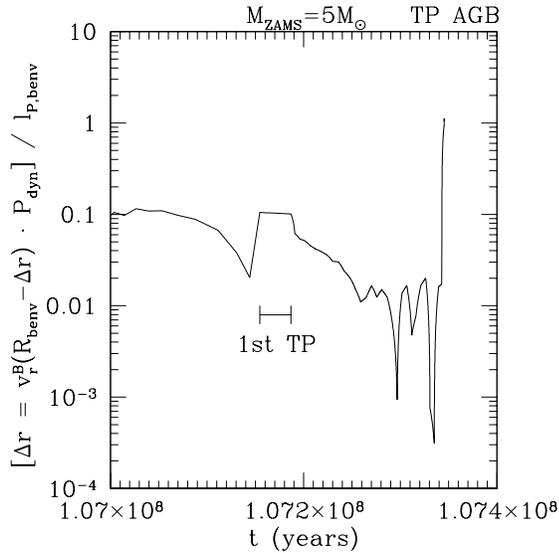}
\epsscale{1.0}
\caption{Thickness $\Delta r$ of magnetized shell below the core-envelope boundary which remains in contact with the inner convection zone
over the dynamo period $P_{\rm dyn}$. See Equation (\ref{eq:Delta_r_buoyancy}).  Here $\Delta r$ is normalized to the local pressure 
scaleheight.  Plot covers the thermally pulsating AGB phase of the $5\,M_\odot$ model, with $B_\phi = (4\pi \rho v_{\rm con}^2)^{1/2}$.}
\vskip .2in
\label{fig:delta_r_TP_AGB}
\end{figure}

Figure \ref{fig:delta_r_TP_AGB} shows the resulting buoyant shell thickness, in comparison with the local scale height,
during the AGB phase of the $5\,M_\odot$ model.
The buoyant shell thickness reaches $\sim 0.1 l_P(R_{\rm benv})$ before and during the first thermal pulse, and then relaxes to a value
about 10 times smaller.


\subsection{Seeding a Toroidal Magnetic Field \\ in the Tachocline}\label{s:bseed}

We now consider the minimal seed field that will trigger dynamo feedback in the tachocline, thereby facilitating the magnetic coupling
of core and envelope.  We suppose that this seed field is generated within the convective envelope, and is deposited in the 
outer radiative core by the mass flow toward the burning shell(s).

This seed field must pass a certain threshold if linear winding in the tachocline is to generate a buoyantly unstable toroidal field.
The time available for linear winding depends on the depth $\Delta r$ below the radiative-convective boundary.  Consider
an unmagnetized core in which matter flowing across the convective boundary sustains an angular velocity profile 
$\Omega = \Omega_{\rm benv}(r/R_{\rm benv})^{-2}$.  Near the boundary, the toroidal magnetic field that is sourced
by a seed radial field $B_{\rm seed}$ is $B_\phi = -2B_{\rm seed}\Omega_{\rm benv}\Delta r/v_r^{\rm nuc}$.  
Substituting this into Equation (\ref{eq:buoy}) and setting $v_r^B > |v_r^{\rm nuc}|$ gives 
\be\label{eq:bseed}
{B_{\rm seed}^2\over 4\pi \rho(R_{\rm benv}) v_{\rm con}^2} > { N^2 \over 2 \Omega^2 }  { P |v_r^{nuc}|^3  \over  v_{con}^2 F_{rad} } \ .
\ee

\begin{figure}[!]
\figurenum{6}
\plotone{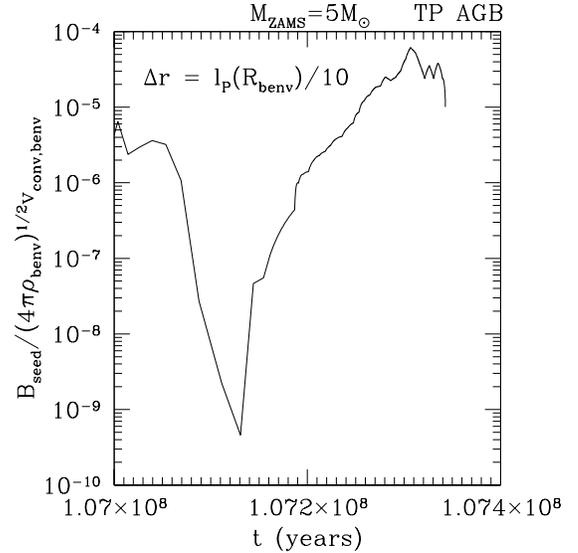}
\vskip 0in
\epsscale{1.0}
\caption{Minimal seed poloidal magnetic field, deposited in the outer core of an AGB star, which will grow by linear winding to a sufficient 
strength that radiatively driven magnetic buoyancy overcomes the downward drift of envelope material to the burning shell(s).
Equation (\ref{eq:bseed}) is evaluated at depth $\Delta r = l_{P,\rm benv}/10$ below core-envelope boundary.
Plot covers same thermally pulsating AGB phase of $5\,M_\odot$ model as Figure \ref{fig:delta_r_TP_AGB}.}
\vskip .2in
\label{fig:B_seed_TAGB}
\end{figure}

This minimal seed field for dynamo feedback turns out to be quite weak.  Figure \ref{fig:B_seed_TAGB} shows 
${B_{\rm seed} / \sqrt{4\pi \rho(R_{\rm benv})} v_{\rm con}}$ during the final AGB evolution of a $5\,M_\odot$ star, measured at a 
distance $\Delta r = l_P(R_{\rm benv})/10$ below the core-envelope boundary.  Much of the C/O mass deposition from double shell burning 
is concentrated around $t \sim 1.0716\times10^8$ yr, when this quantity is as small as $\sim 10^{-9}$; it rises to $\sim 10^{-4}$ 
during the phase of rapid thermal pulses.

Hydrodynamic turbulence is strongly suppressed in the tachocline.  The mean-field approach to dynamo action then
suggests (see the review in Section \ref{s:dynthresh}) that dynamo activity should be easily sustained in the
tachocline as long as the toroidal magnetic field can be rotated back into the poloidal direction.  This effect is present in numerical
simulations of a shear layer \citep{ClinBC2003}; and is conjectured to operate in a more global dynamo model where
toroidal magnetic field formed in the tachocline diffuses into the convective envelope \citep{park1993}.
Estimates of the equipartition toroidal magnetic field and the large-scale Maxwell stress are given by equations 
(\ref{eq:bphieq}) and (\ref{eq:reyn}).

\begin{figure}[!]
\figurenum{7}
\plotone{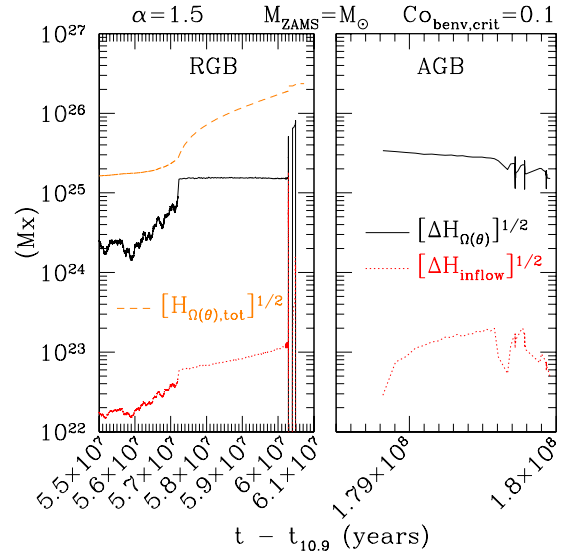}
\vskip 0in
\epsscale{1.0}
\caption{The magnetic helicity (expressed as a magnetic flux $\Delta {\cal H}^{1/2}$) that 
accumulates in the time $t_{\rm drift}
= \ell_P/|v_r^{\rm nuc}|$.  RGB (left) and AGB (right) evolution of $M_{\rm ZAMS} = 1\,M_\odot$ model
with a Jupiter companion in an initial orbit $a_i = 1$ AU. 
Most of the helicity generated on the RGB is concentrated near the tip (orange short-dash curve shows cumulative helicity).
This gives $M_{\rm benv}(5.65\times10^7)$ $\sim 0.388M_\odot$ as the 
effective base of the magnetized region.  The net mass passing through the dynamo-active layer 
is then $\sim0.12M_{\odot}$ on the RGB, and $\sim 0.014 M_{\odot}$ on the AGB. 
The greater helicity accumulated through the `$\Omega(\theta)$' channel (Equation (\ref{eq:helicityf})), 
as compared with the `inflow' channel (Equation (\ref{eq:helicityd})), is reflected in the final WD magnetic 
field (Figure \ref{fig:B_WD_three_dynamos_vs_L_orb}).}
\vskip .2in
\label{fig:Delta_H_sqrt_TP-AGB_solar}
\end{figure}

\begin{figure}[!]
\figurenum{8}
\plotone{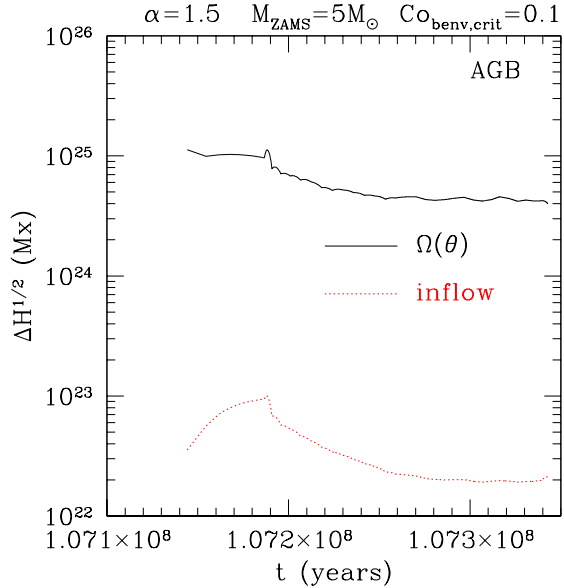}
\epsscale{1.0}
\caption{Comparison of magnetic helicity generated over $t_{\rm drift} = \ell_P/|v_r^{\rm nuc}|$ 
through the same two channels as Figure \ref{fig:Delta_H_sqrt_TP-AGB_solar}, but now in the
$M_{\rm ZAMS} = 5\,M_\odot$ model without a planetary companion.
Only the AGB contributes directly to final WD magnetization in this case, within
a mass shell $\sim 0.044\,M_\odot$.  The largest net contribution comes from the first thermal pulse.}
\vskip .2in
\label{fig:Delta_H_sqrt_TP-AGB_5M_solar}
\end{figure}

\section{White Dwarf Magnetic Field} \label{s:finalB}

After magnetic helicity accumulates in the core, we allow the magnetic field to relax to a more isotropic configuration.  
A first estimate of the flux threading one hemisphere is
\be\label{eq:phicore}
\Phi_{\rm core} \simeq {\cal H}^{1/2}.
\ee
The final poloidal magnetic field of a WD of radius $R_{\rm wd}$ is
\be\label{eq:bwd}
B_{\rm wd} \sim \frac{\Phi_{\rm core}}{\pi R_{\rm wd}^2}.
\ee

We first evaluate the magnetic helicity that is deposited in a post-AGB star,
using the $1\,M_\odot$ and $5\,M_\odot$ stellar models constructed using MESA.
This stored helicity is then converted to a dipolar magnetic
field in the WD remnant using Equations (\ref{eq:phicore}) and (\ref{eq:bwd}).  

This estimate presupposes
that toroidal and poloidal magnetic fluxes evolve to an energy minimum with $\Phi_p \sim \Phi_\phi$, all the while preserving
the product helicity ${\cal H} \sim \Phi_p\Phi_\phi$.  Such an interchange may occur later in the life of a WD as the
result of ohmic diffusion (Section \ref{s:B_field_emergence_decay}); or earlier on as the result of relaxation within
episodes of confined convection (such as the core helium flash, or helium shell flashes).

\subsection{Accumulation of Magnetic Helicity in a \\ Radiative Core Relaxing to Solid Rotation} \label{s:mass_drift_dynamo}

The zone of interest here is the tenuous mantle of radiative material that is sandwiched between the hydrogen-depleted 
core and the base of the convective envelope.  Here only a small mass $\lesssim 10^{-3}\,M_\odot$ lies within
a scaleheight, and so we model the mass inflow through the mantle as quasi-steady,
\be
\dot M_{\rm core} = 4\pi r^2 |v_r| \rho(r).
\ee 

The poloidal magnetic field is approximated as radial
with flux function (\ref{eq:flux}), and the toroidal magnetic field is expressed in terms of a radial magnetic twist 
as $B_\phi(r,\theta) = (\partial\phi_B/\partial r) r\sin\theta B_r$.  Then helicity accumulates in the core at the rate
\be\label{eq:helicityf}
{d{\cal H}\over dt} = -2\pi\delta v_r\int d\theta \Phi_r {\partial \Phi_r\over\partial\theta} {\partial\phi_B\over \partial r},
\ee
where $\delta v_r$ is the flow velocity relative to the core-envelope boundary.  

Now consider the torque imparted to a mass shell by the Maxwell stress $B_rB_\phi/4\pi$,
\be
{d\over dt}\left(\delta m {2\over 3}r^2\bar\Omega\right) = \delta\left[\int dS\; r\sin\theta {B_rB_\phi\over 4\pi}\right].
\ee
Here the Lagrangian time derivative follows the radial flow at speed 
$v_r$. Since $d(\delta m)/dt = 0$, one has, following Equation (\ref{eq:flux}),
\be \label{eq:torque}
{d\over dt}\left({2\over 3}r^2\bar\Omega\right) = {1\over 8\pi \rho r^2} 
\int d\theta \sin\theta \left({\partial \Phi_r\over\partial \theta}\right)^2 {\partial^2\phi_B\over \partial r^2}.
\ee
Both hemispheres generally contribute in the same sense to the torque.

\subsubsection{Helicity Sourced by Radial Inflow}\label{s:massdrift}

We suppose that the magnetic field is strong enough that nearly solid rotation is maintained in the radiative layers, 
$|\partial\Omega/\partial r| \ll \bar\Omega/r$. Each mass shell starts with a mean angular frequency $\Omega_{\rm benv}$ that is imposed by 
the lower part of the convective envelope. The twist that is required to maintain a weak angular velocity gradient is found from Equation 
(\ref{eq:torque}),
\be \label{eq:twistb}
{\partial^2\phi_B\over\partial r^2} \sim -{2\dot M_{\rm core}\Omega_{\rm benv}r \over (B_r r^2)^2},
\ee
where the coefficient on the right-hand side corresponds to $B_r$ independent of $\theta$. From Equation (\ref{eq:phiB}) one has
\be
{\partial\phi_B\over\partial r} \sim {r\over |v_r|}{\partial\Omega\over\partial r}.
\ee
Nearly solid rotation is maintained if
\be \label{eq:solid}
v_{A,r}^2 \equiv {B_r^2\over 4\pi \rho} \gg v_r^2.
\ee

Now consider the magnetic helicity that is advected into the growing radiative core. We evaluate the magnetic twist in Equation 
(\ref{eq:helicityf}) from Equation (\ref{eq:twistb}), obtaining
\be \label{eq:helicityd}
{d{\cal H}\over dt} = f(\Delta M_{\rm core})\cdot \pi \dot M_{\rm core} \Omega_{\rm benv} R_{\rm benv} ^2 \delta v_r(R_{\rm benv} )
\ee
at $r = R_{\rm benv}$.

The factor $f$, with indeterminate sign but magnitude $|f| < 1$, represents the cancellation between hemispheres. 
We discuss its origin in Section \ref{sec:cancellation_factor}.

\subsubsection{Magnetic Twist Compensating a Latitude-dependent Convective Torque}\label{s:contorque}

A greater Maxwell stress results from the application of convective stresses to the outer radiative core.  
A convection zone generally sustains latitudinal gradients in rotation.  Part of the differential
Reynolds stress is transmitted through the tachocline to the core by a dynamo-generated magnetic
field (Equation (\ref{eq:reyn})).  As a working estimate, we use
\be\label{eq:maxcon}
{B_rB_\phi\over 4\pi}  = \varepsilon_B \rho(R_{\rm benv}) (\Omega_{\rm benv} R_{\rm benv})^2,
\ee
with $\varepsilon_B \sim 10^{-3}$, corresponding to a dynamo period $P_{\rm dyn} \sim 10
[\Delta\Omega(R_{\rm benv})]^{-1}$ and a pole-equator offset $\Delta\Omega(R_{\rm benv}) \sim \Omega_{\rm benv}$.  

Given that the deeper parts of the radiative core are coupled magnetically to the tachocline,
uniform rotation can be maintained in the outer core only if this stress is divergence-free below the tachocline.
The corresponding magnetic helicity flow into the core through the advection of this twisted field is
\be \label{eq:helicitye}
{d{\cal H}\over dt} = f(\Delta M_{\rm core})\cdot \pi \varepsilon_B \dot M_{\rm core} R_{\rm benv} (\Omega_{\rm benv} R_{\rm benv})^2.
\ee

The contributions to ${\cal H}$ from equations (\ref{eq:helicityd}) and (\ref{eq:helicitye}) are labelled
in the Figures as `inflow' and `$\Omega(\theta)$', respectively.

\subsubsection{Cancellation Between Hemisphere} \label{sec:cancellation_factor}

One feature of Equations (\ref{eq:helicityd}) and (\ref{eq:helicitye}) stands out:  the growth of ${\cal H}$ is
independent of the seed radial magnetic field, as long as it is strong enough to satisfy the inequality (\ref{eq:solid}).  
This means that the contribution to ${\bf A}\cdot{\bf B}$ {\it in one hemisphere} is not strongly affected by fluctuations 
in the magnitude and sign of the poloidal magnetic field that is supplied to the radiative layer by a dynamo operating near 
the convective boundary.

A net mass $\Delta M_{\rm core}$ flows through the convective-radiative boundary over the duration of dynamo activity.  In the case of 
an AGB star, this is approximately the increase in the mass of the C/O core due to double-shell burning.  Note that $\Delta M_{\rm core}$
is generally much larger than the mass of the rarefied core mantle ($\sim 10^{-3}\,M_\odot$).
The low mass of the mantle imposes a short-timescale cutoff to fluctuations in the summed ${\cal H}$ from the two 
hemispheres.  We express this cutoff in terms of the mass within a scale height below the core-envelope boundary,
\be
\Delta M_{\rm min} \sim {4\pi Pr^4\over GM_{\rm core}}\biggr|_{R_{\rm benv} }.
\ee
Then we take
\be\label{eq:fM}
f(\Delta M_{\rm core}) \sim \left({\Delta M_{\rm min}\over \Delta M_{\rm core}}\right)^{1/2}.
\ee

\subsection{Results}

First consider the $M_{\rm ZAMS} = 1\,M_\odot$ model.   The angular momentum profile of the star is evolved
in post-processing using the method of Section \ref{s:rotation} and Paper I.  Companion planets with a range of
masses (Earth, Neptune, and Jupiter) are placed at an initial semi-major axis $a_i = [0.5,0.75,1,1.5,2]$ AU and with a range of
orbital eccentricities.

Magnetic fields generated during the RGB and AGB phases both contribute to the 
remnant WD field, but at different depths in the star (see Figure \ref{fig:M_star_M_base_M_He_core_M_CO_core_TRGB_solar}). A net mass 
$\sim 0.135M_{\odot}$ passes through the core-envelope boundary while conditions are favorable for magnetic field growth.
\footnote{For example, we do not include the first dredge-up phase, during which the convective envelope grows in mass.}

\begin{figure*}[!]
\figurenum{9}
\epsscale{1.0}
\plottwo{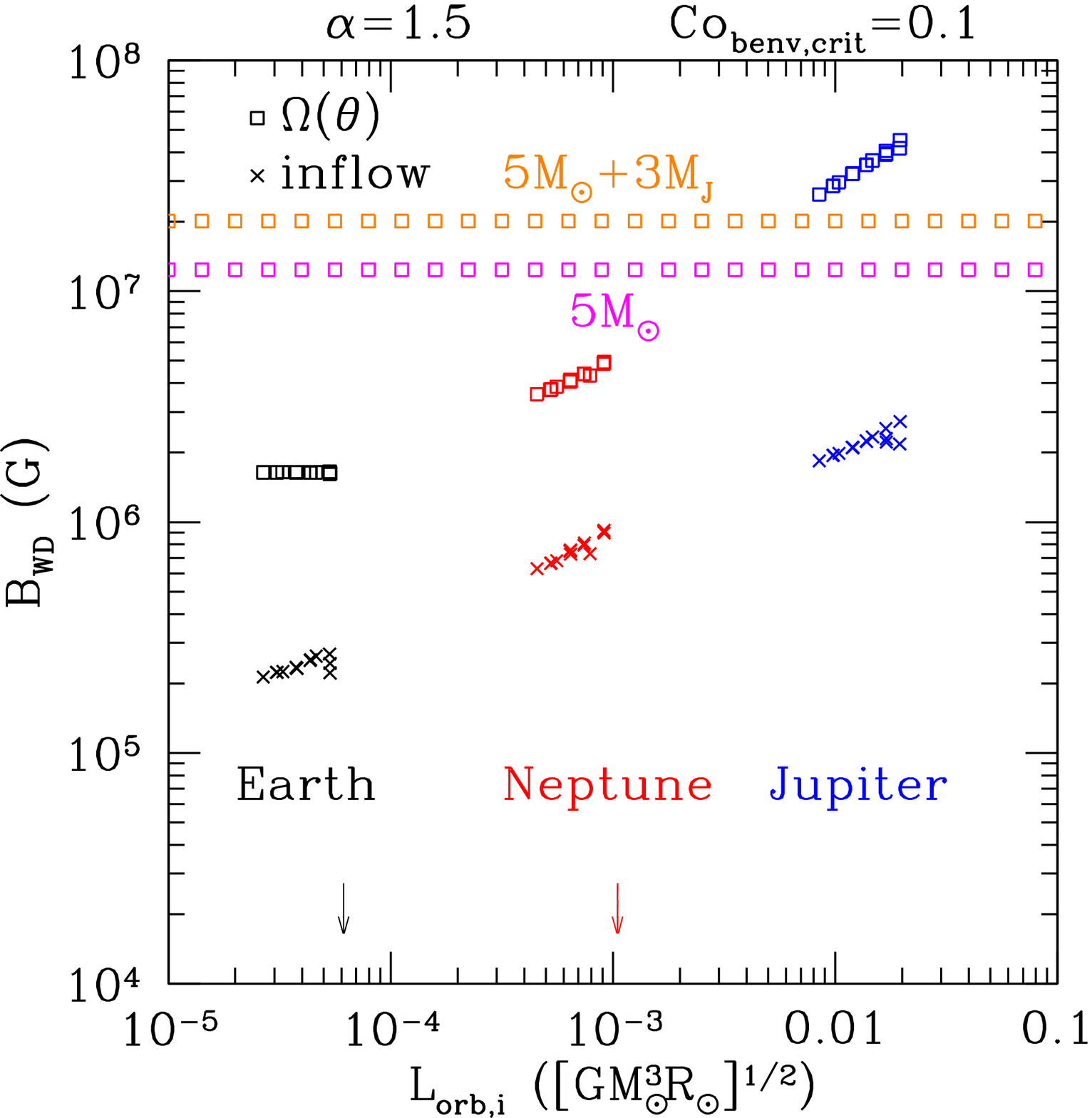}{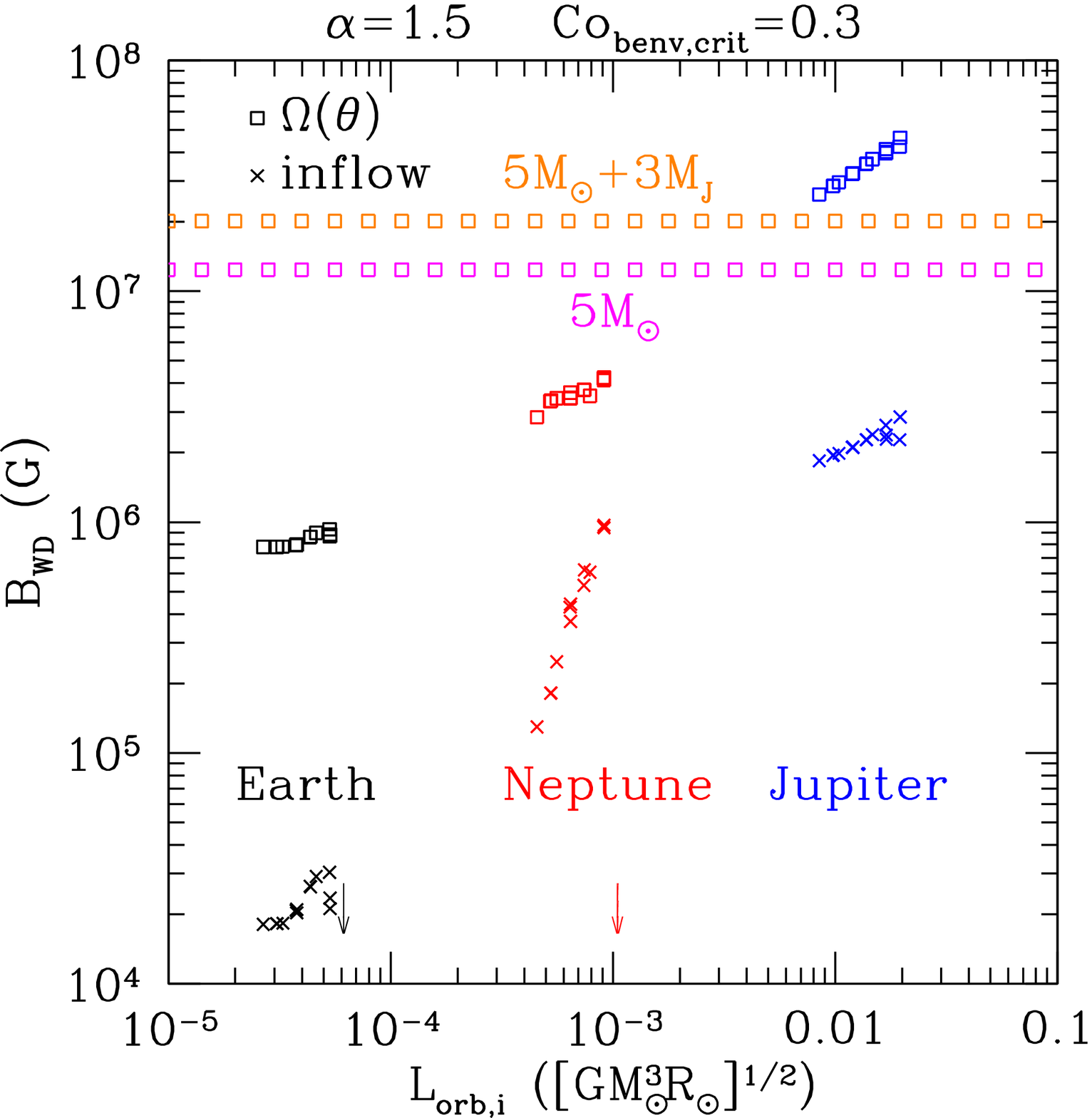}
\vskip 0in
\caption{Dipole magnetic field, Equation (\ref{eq:bwd}), left behind in the WD remnant of a $1\,M_\odot$ star
interacting with a planet (black = Earth mass, red = Neptune mass, blue = Jupiter mass).  
Result is plotted separately for the two helicity channels described in the text. Horizontal axis shows 
initial orbital angular momentum of planet companion. Each tick represents an average over realizations of
a given set of orbital initial conditions ($a_i$ and $e_i$). Magenta squares show result for
$5\,M_\odot$ progenitor with initial equatorial rotation speed 50 km s$^{-1}$ and no companion; orange squares same
model with angular momentum of $3M_J$, $a_i = 2$ AU planet added during early AGB. 
Dynamo shuts off when ${\rm Co}_{\rm benv} < {\rm Co}_{\rm benv,crit} = 0.1$ (left panel) or ${\rm Co}_{\rm benv,crit} = 0.3$ (right panel).}
\vskip .2in
\label{fig:B_WD_three_dynamos_vs_L_orb}
\end{figure*}

Figure \ref{fig:Delta_H_sqrt_TP-AGB_solar} shows the helicity that accumulates in the brief period $t_{\rm drift} = \ell_P/|v_r^{\rm nuc}|$ 
when stellar material settles through the outer scale height of the radiative core. The dominant contribution comes from the 
`$\Omega(\theta)$' channel.  

In the $5M_{\odot}$ star, only helicity accumulated on the AGB contributes directly to the final WD magnetic field
(Figure \ref{fig:M_star_M_base_M_He_core_M_CO_core_TRGB_5M_o}).  During this phase $\sim 0.044 M_{\odot}$ passes 
through the active dynamo region (given our normalization $\eta_B = 0.05$ of the Bl\"{o}cker mass loss formula). 
We see in Figure \ref{fig:Delta_H_sqrt_TP-AGB_5M_solar} that the
initial phase of the dynamo, which occurs during the first thermal pulse, contributes the majority of ${\cal H}$.
In the 5 $M_{\odot}$ model, we simplified the treatment of the thermal pulses by linearly interpolating
all of the relevant quantities between the local minima in $R_{\rm benv}$ after each He shell flash, so as to focus on the 
mean growth of core mass and ${\cal H}$.   The stellar model considered in Figure \ref{fig:Delta_H_sqrt_TP-AGB_5M_solar}
has no planetary companion.

The resulting WD dipolar field is shown in Figure \ref{fig:B_WD_three_dynamos_vs_L_orb}
for both the $1\,M_\odot$ and $5\,M_\odot$ models, using
the mass-radius relation of \cite{zapos1969}.
A strong magnetic field in the $0.55\,M_\odot$ WD remnant of the $1\,M_\odot$ progenitor depends on the 
injection of angular momentum from a planetary companion (as considered here), or a tidal interaction with a stellar
companion.   We find that $B_{\rm wd}$  is an increasing function of the angular momentum absorbed;
the various points correspond to different realizations of the orbital evolution.  

Dipole fields approaching $10^8$ G are achievable through the `$\Omega(\theta)$' channel, following the absorption of a Jupiter-mass planet.
Note that an Earth-mass planet does not signicantly augment the angular momentum left over at the end of the MS spindown, and so
this case corresponds closely to an isolated star that does not interact with planets. 

Figure \ref{fig:B_WD_three_dynamos_vs_L_orb} focuses on those runs which end in the engulfment of the planet. However there are cases in which
the planet is pushed out into a large orbit due to the convective quadrupole. The planet therefore gains angular momentum from the star, which
in turn causes the star to spin in the opposite direction. In some cases this leads to the development of a dynamo as well.  We ignore these
particular cases.


An external source of angular momentum injection is not required to sustain a dynamo in the $5\,M_\odot$ progenitor.  
Nonetheless, $B_{\rm wd}$ depends on the C/O mass that accumulates during double shell burning, and therefore on the 
mass-loss rate during the expulsion of the envelope.  

\begin{figure}[!]
\figurenum{10}
\epsscale{1.0}
\plotone{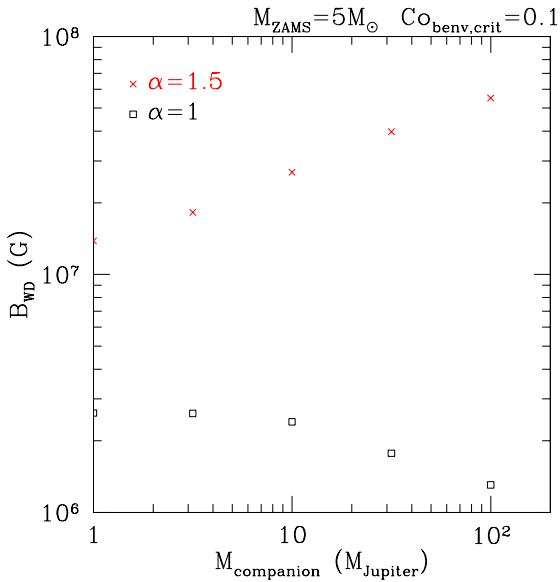}
\caption{Dipole magnetic field obtained via the $\Omega(\theta)$ channel in the $0.87\,M_\odot$ remnant of a $5\,M_\odot$ star that
absorbs an orbiting companion ($a_i = 2$AU) of various masses during the early AGB phase ($R_\star = 200R_\odot$). 
Top points: rotation profile $\Omega(r) \propto r^{-1.5}$ ($r^{-2}$) in the inner (outer) convective envelope 
($\alpha = 1.5$ in Equation (\ref{eq:ominner})).  
Bottom points: $\alpha = 1$.}
\vskip .2in
\label{fig:B_WD_high_mass}
\end{figure}

The effect of adding a relatively massive companion to the $5\,M_\odot$ model is considered in Figure
\ref{fig:B_WD_high_mass}.  The absorption of a $3\,M_J$ planet significantly augments the remnant dipole 
field in the $0.87\,M_\odot$ WD. We see a steady increase in the remnant field as the companion mass increases to the brown 
dwarf range, or even to $\sim 0.1\,M_\odot$, for the angular velocity profile considered here ($\alpha = 3/2$). 
For comparison we show the magnetic fields obtained by a shallower inner rotation profile ($\alpha = 1$), which may be applicable
depending on the strength of the Coriolis back reaction. In this case an increase in angular momentum results in a decrease of remnant 
magnetic field, caused by a decrease in ${\rm Co_{benv}}$ due to the shallow $\tau_{\rm conv}(r)$ profile.

\section{White Dwarf Rotation} \label{s:WD_rotation}

We now analyze the spin evolution of the core and envelope of our stellar models.
Our focus is on the magnetic {\it de}-coupling between the core and envelope.   Three factors determine the
effectiveness with which a poloidal magnetic field transfers angular momentum across the core-envelope boundary,
and establishes solid rotation in the outer core:  i) the timescale over which mass is exchanged between
core and envelope; ii) the direction of this exchange; and iii) the Coriolis parameter in the inner envelope.

The internal structure of the star evolves slowly enough during the early ascent of the giant branches, and during
core helium burning, that a magnetic coupling between core and envelope is very difficult to avoid
(see Figure 2 of Paper I).  There is a much more rapid exchange of mass between envelope and core 
near the tips of the RGB and AGB.   The first giant phase is (except in cases of close binary interaction)
followed by more extended evolution.  So our focus here is on the final AGB superwind phase during which most
of the envelope mass is removed.  

The ability of the inner envelope to generate a magnetic field is most relevant 
while the radiative core is growing in mass.  Then magnetized material is advected downward into the outer
core, where it can communicate changes in rotation rate in the envelope to the inner core.  The core 
grows in mass during the thermally pulsating AGB phase, after averaging over the pulsations;  but following
the contraction of the envelope the superwind causes a small but rotationally significant decrease in mass.

An important effect involves the re-expansion of the envelope following a late thermal pulse:
following this about $10^{-3}\,M_\odot$ of hydrogen-rich material is transfered 
to the convective envelope.  If core and envelope are still magnetically coupled, this 
causes a dramatic spindown of the core through the internal exchange of angular momentum.

Faster WD spin is mediated by an early loss of angular momentum from 
the surface of an AGB star.  This loss, if strong enough, will push the Coriolis parameter ${\rm Co}_{\rm benv}$ of 
the inner envelope below unity well before the envelope contracts.  The remnant WD spin period depends on the critical value of 
${\rm Co_{ benv}}$ below which the dynamo effectively shuts off. 
If the transition to ${\rm Co}_{\rm benv} < {\rm Co}_{\rm benv,crit}$ takes place before the final shell flash, then the angular momentum of 
the core is effectively frozen in and inherited by the WD.  On the other hand,
if this transition occurs afterward, then we expect that the rotation of the outer core remains coupled to the inner
envelope as the photosphere of the star contracts.

\begin{figure}[!]
\figurenum{11}
\epsscale{1.0}
\plotone{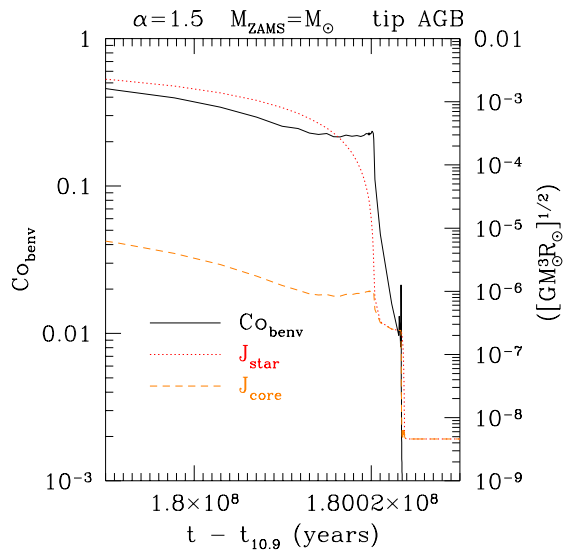}
\vskip 0in
\caption{Rotational angular momentum of $M_{\rm ZAMS} = 1\,M_\odot$ AGB star (dotted red line) and its core (short-dash orange line) just 
before and during the super-wind phase. The star has absorbed a Jupiter-mass companion near the tip of the RGB. 
Black line: Coriolis parameter at base of convective envelope. Rotation profile
$\Omega(r) \propto r^{-1.5}$ ($r^{-2}$) in the inner (outer) convective envelope.  
When calculating remnant WD spins, we freeze $\Omega_{\rm core}$ when and if ${\rm Co}_{\rm benv}$ drops below a critical value, 
beyond which time the curves plotted here do not apply.}
\vskip .2in
\label{fig:Co_base_and_J_star_J_core_TAGB_closer_solar}
\end{figure}

\begin{figure}[!]
\figurenum{12}
\epsscale{1.0}
\plotone{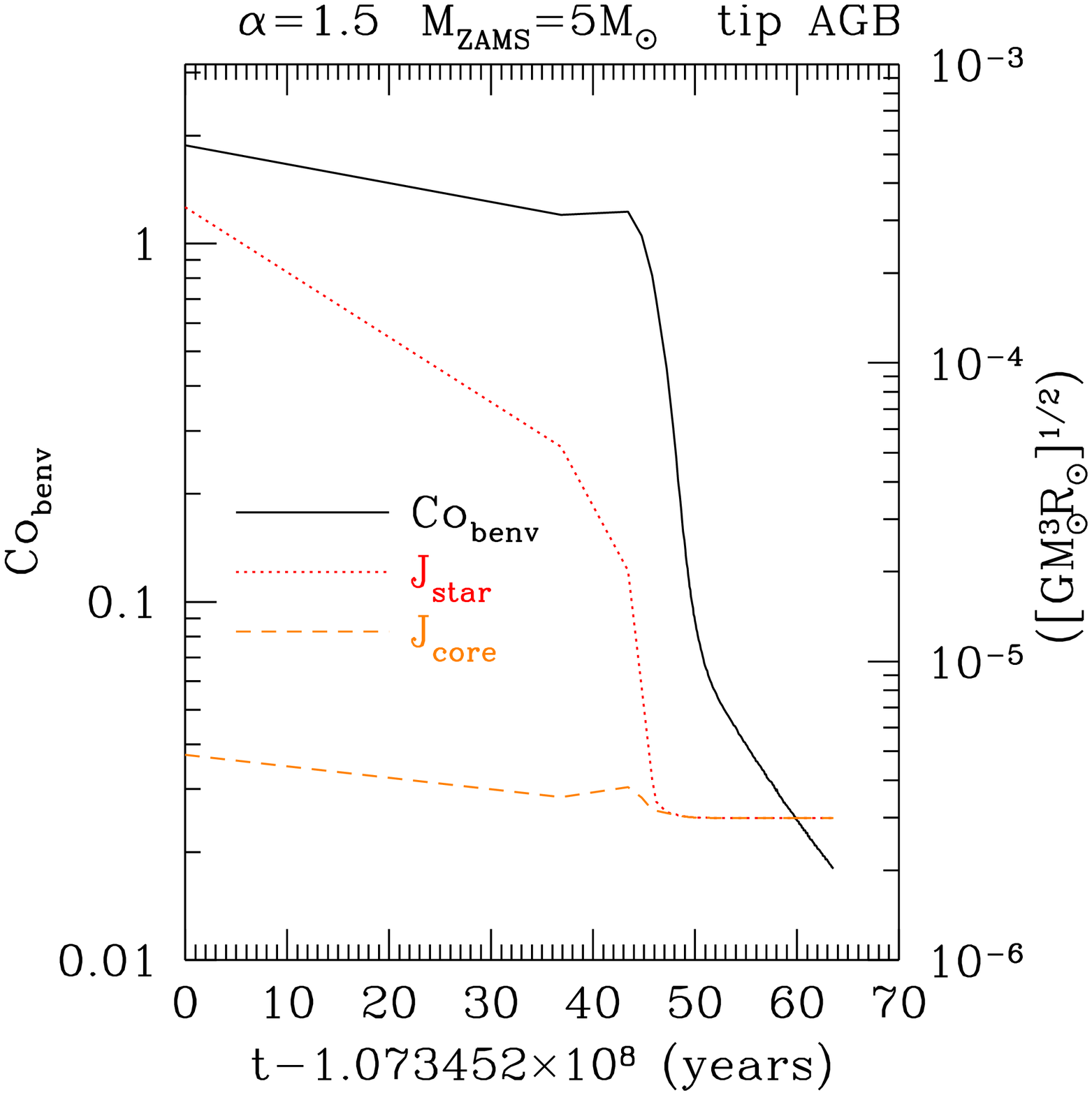}
\vskip 0in
\caption{Similar to Figure \ref{fig:Co_base_and_J_star_J_core_TAGB_closer_solar}, but for the $M_{\rm ZAMS} = 5M_{\odot}$ star.}
\vskip .2in
\label{fig:Co_base_and_J_TAGB_closer_5M_solar_time_shift}
\end{figure}

Figures \ref{fig:Co_base_and_J_star_J_core_TAGB_closer_solar} and \ref{fig:Co_base_and_J_TAGB_closer_5M_solar_time_shift} show 
the evolution of the net stellar angular momentum $J_\star$ and the core angular momentum $J_{\rm core}$ during the AGB phase, under the assumption 
of a tight core-envelope coupling.   The internal rotation profile, the prescription for mass and angular momentum loss,
and the interaction with a planetary companion, are the same as those described in Sections \ref{s:intro} \& \ref{s:basic}
and detailed in Paper I.  As the star experiences strong mass loss, there is a rapid drop in $J_\star$.

During this phase, $J_\star$ is still dominated by the envelope.  
Also plotted is ${\rm Co}_{\rm benv}$ for comparison.  Core-envelope decoupling becomes plausible at 
${\rm Co}_{\rm benv} = {\rm Co}_{\rm benv,crit} \sim 0.1$-0.3, given the strong dependence of magnetic field
on rotation that is observed in deeply convective MS stars \citep{reinbb2009}.  

Whether the transition to ${\rm Co}_{\rm benv} < {\rm Co}_{\rm benv,crit}$ is encountered before the final
He shell flash depends on both $J_\star$ and the value of ${\rm Co}_{\rm benv,crit}$.  The transition occurs earlier for a larger critical 
Coriolis parameter, as in our $1\,M_\odot$ model with ${\rm Co}_{\rm benv} = 0.3$.
But we find a tighter restriction on $J_\star$ when ${\rm Co}_{\rm benv}$ is reduced to 0.1:  
then the absorption of a Jupiter will cause core and envelope to remain coupled past the final flash
in the $1\,M_\odot$ model, but the absorption of a Neptune (or of no planet) allows a transition to a weakly magnetized envelope.  
We find that ${\rm Co}_{\rm benv}$ generally remains above ${\rm Co}_{\rm benv,crit}$ at the final shell flash in the $5\,M_\odot$ model, 
implying a tight core-envelope coupling  during the transition to a post-AGB star.

In the following two sections, we separately consider the rotational evolution of our stellar models with, and without,
a transition to weak core-envelope coupling.

\subsection{Dependence of White Dwarf Spin on \\ Coriolis Parameter for Core-Envelope Decoupling} \label{s:Co_base_critical}

\begin{figure}[!]
\figurenum{13}
\epsscale{1.0}
\plotone{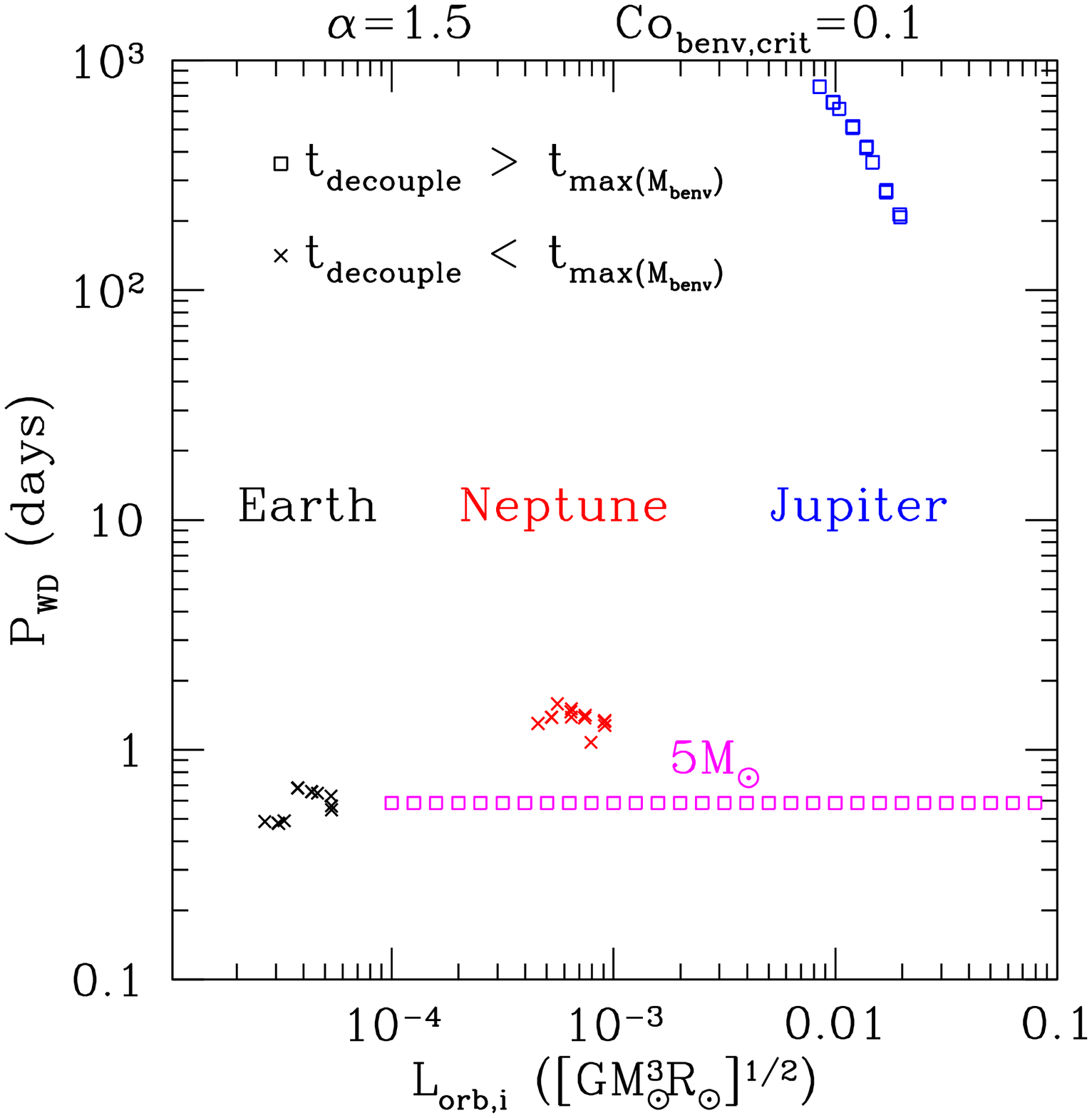}
\vskip 0.2in
\plotone{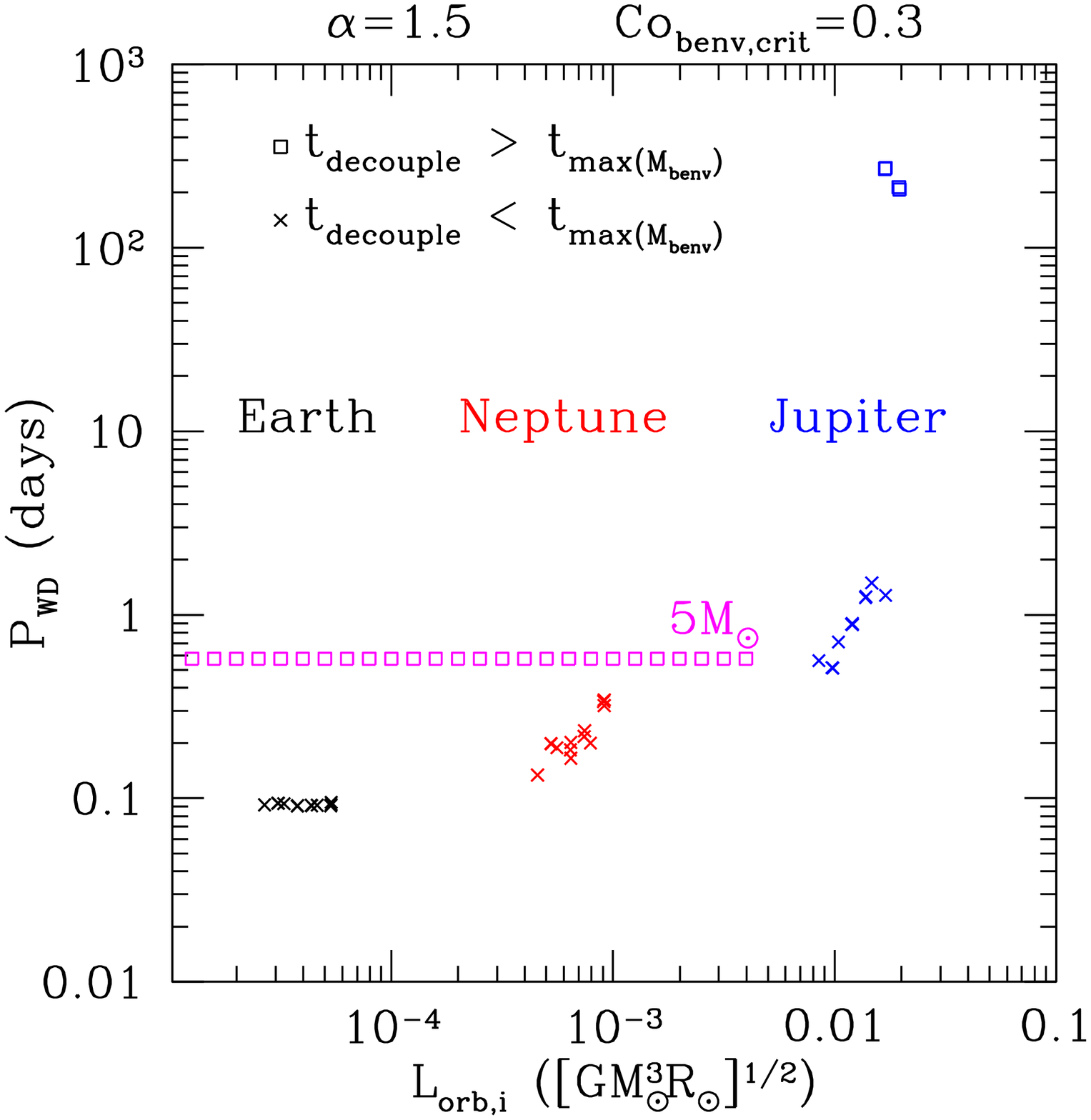}
\vskip 0in
\caption{Rotation period of $0.55\,M_\odot$ WD remnant of $1\,M_\odot$ star, as a function of the initial orbital 
angular momentum of planetary companion. Squares: Core and envelope remain magnetically coupled until the post-AGB phase.  
x's:  Core-envelope decoupling before the final helium shell flash, with $J_{\rm wd}$ set to  $J_{\rm core}$ at decoupling.    
Decoupling takes place when ${\rm Co}_{\rm benv}$ drops below ${\rm Co}_{\rm benv,crit} = 0.1$ (upper panel) or
$0.3$ (lower panel). Horizontal magenta line: result for $0.87\,M_\odot$ remnant of $M_{\rm ZAMS} = 5\,M_\odot$ star with 
50 km s$^{-1}$ equatorial rotation period.}
\vskip .2in
\label{fig:P_rot_WD_Co_crit_w_5M_o}
\end{figure}

The WD spin period that results from core-envelope decoupling at ${\rm Co}_{\rm benv,crit} = 0.1$ and 0.3 in the $1\,M_\odot$
model is shown in Figure \ref{fig:P_rot_WD_Co_crit_w_5M_o}.  
For the more conservative estimate ${\rm Co}_{\rm benv,crit} = 0.1$, we find that $P_{\rm wd}$ ranges over 
$0.5$-$1.5$ d.

In this situation, where the timing of the transition to weak magnetization in the envelope depends on $J_\star$,
$P_{\rm wd}$ depends on the initial orbital angular momentum of the planetary companion.  A lower total angular momentum results in faster 
decoupling, and therefore a {\it larger} trapped core angular momentum:  the core rotation frequency at decoupling is tied to the convective 
timescale in the inner envelope, and to the structure of the star.

\begin{figure}[!]
\figurenum{14}
\epsscale{1.0}
\plotone{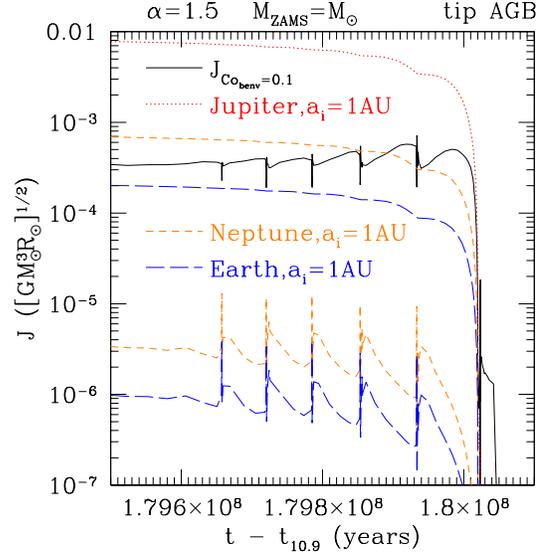}
\vskip 0in
\caption{Black solid line: minimum $J_\star$ that maintains ${\rm Co}_{\rm benv} = 0.1$ near the tip of the AGB of a 
$M_{\rm ZAMS} = 1\,M_\odot$ star. For comparison, colored lines show calculated angular momentum of star 
absorbing a Jupiter (red dotted), Neptune (orange short-dashed) or Earth (blue long-dashed) starting with $a_i=1$ AU, $e_i=0$. 
$J_{\rm core}$ is shown separately for Neptune and Earth interaction, where core and envelope decouple relatively early.
Decoupling is delayed to the final super-wind phase when the companion has a Jupiter mass,
resulting in lower $J_{\rm core}$ and the longer WD spin seen in Figure \ref{fig:P_rot_WD_Co_crit_w_5M_o}.}
\vskip .2in
\label{fig:J_Co_base_eq_1_TAGB}
\end{figure}

This behavior is explained in Figure \ref{fig:J_Co_base_eq_1_TAGB}, which shows the minimum angular momentum required to 
maintain ${\rm Co}_{\rm benv} = 0.1$ in the $1\,M_\odot$ model near the tip of the AGB.  This is compared with the actual angular momentum
evolution that results from the absorption of a Jupiter, Neptune or Earth, initially orbiting at $a_i=1$ AU with eccentricity $e_i=0$.   

\subsection{White Dwarf Spin in Case of Continued Core-Envelope Coupling}

We observe a significant difference in the remnant angular momentum in two models which maintain a tight 
core-envelope coupling through to the post-AGB phase:  the $5\,M_\odot$ model, and the $1\,M_\odot$ model
which absorbs a Jupiter-mass planet.  In the first case,
downward angular momentum pumping in the convective envelope allows $J_{\rm core}$ to remain relatively
constant during the ejection of the envelope, and only modest spindown is caused by a wind from
the surface of the post-AGB star (Figure \ref{fig:Co_base_and_J_TAGB_closer_5M_solar_time_shift}). 
The spin of the WD remnant is directly 
proportional to the initial spin period of the star, except in the case of the absorption of
a relatively massive ($\gtrsim 3M_J$) companion.  
An equatorial rotation speed $50$ km s$^{-1}$ on the zero-age MS results in $P_{\rm wd} \simeq 0.6$ days.

The initial contraction of the envelope also does not cause much of a reduction in $J_{\rm core}$ in the 
$1\,M_\odot$ model; but we observe a much stronger decrease following the re-expansion and re-contraction
of the envelope that is triggered by a late helium shell instability (Figure \ref{fig:Co_base_and_J_star_J_core_TAGB_closer_solar}).  
This means that the WD remnant of the $5\,M_\odot$ star has a spin period $P_{\rm wd} \sim 15\,$ hr, as compared with $\sim 2$ yr for the 
remnant of the $1\,M_\odot$ star.

To understand how such a dramatic difference in final WD spin could arise, we consider a simplified analytic model
of mass and angular momentum loss from an AGB star.  The envelope of the model star has uniform specific angular momentum,
the core has a fixed radius $R_{\rm benv}$, and the core rotates as a solid body with angular velocity
$\Omega_{\rm core} = \Omega_{\rm benv}$.   Mass is lost from the outer boundary of the star, and is also
exchanged between core and envelope,
\be
{dM_{\rm env}\over dt} = {dM_\star\over dt} - {dM_{\rm benv}\over dt}.
\ee
Here $M_{\rm env}$ is the mass of the convective H-rich envelope, and $M_{\rm benv}$
is the mass of (mainly radiative) material inside the base of the envelope.
The moment of inertia of the core is initially a small fraction of the total effective moment of inertia
\be
I_{\rm eff}  = {J_\star\over \Omega_{\rm benv}} = {2\over 3}M_{\rm env}R_{\rm benv}^2 + I_{\rm core},
\ee
where we parameterize  $I_{\rm core} = (2\epsilon_{\rm core}/3) M_{\rm benv} R_{\rm benv}^2$.   

\begin{figure}[!]
\figurenum{15}
\epsscale{1.0}
\plotone{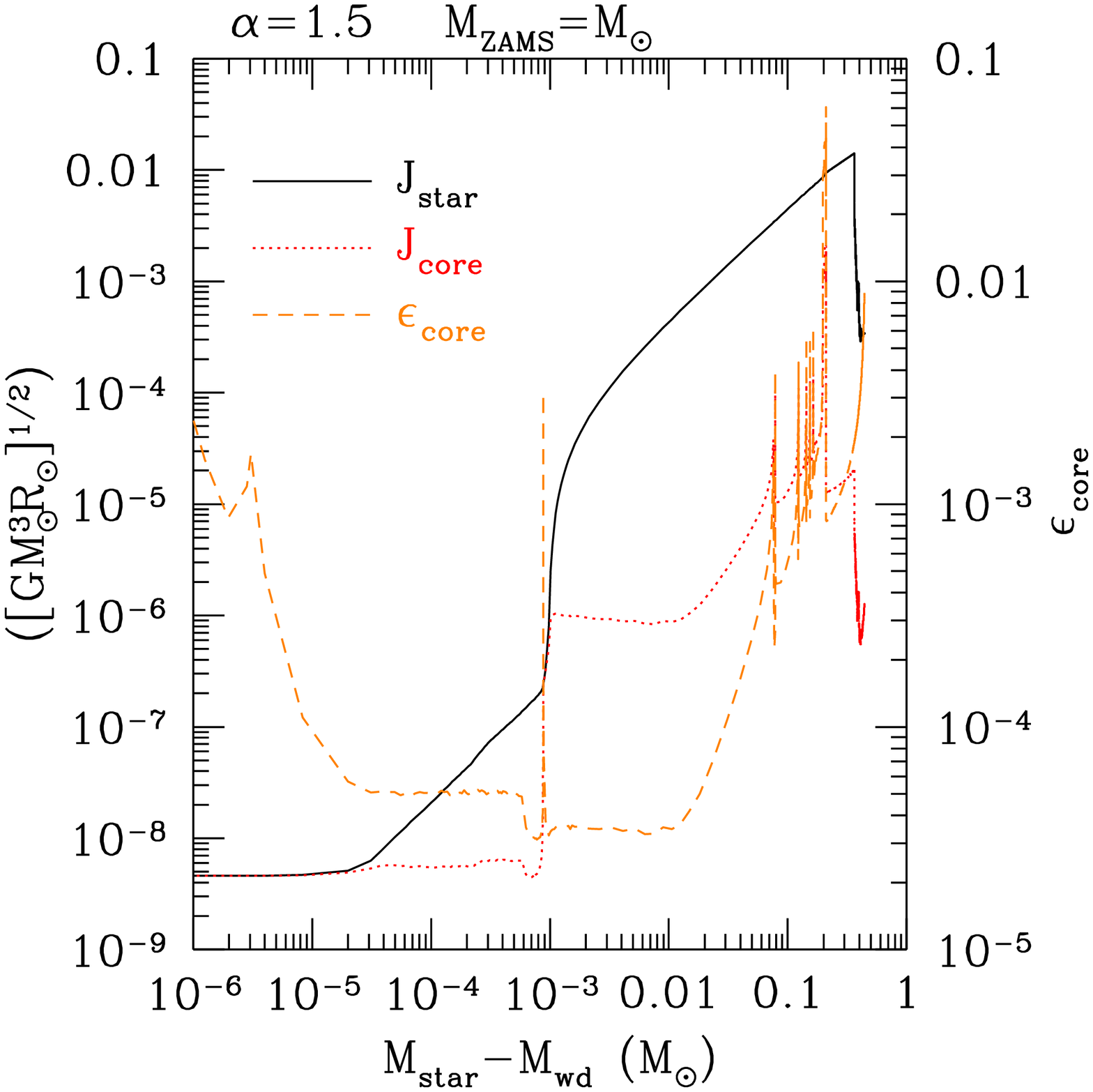}
\vskip .2in
\plotone{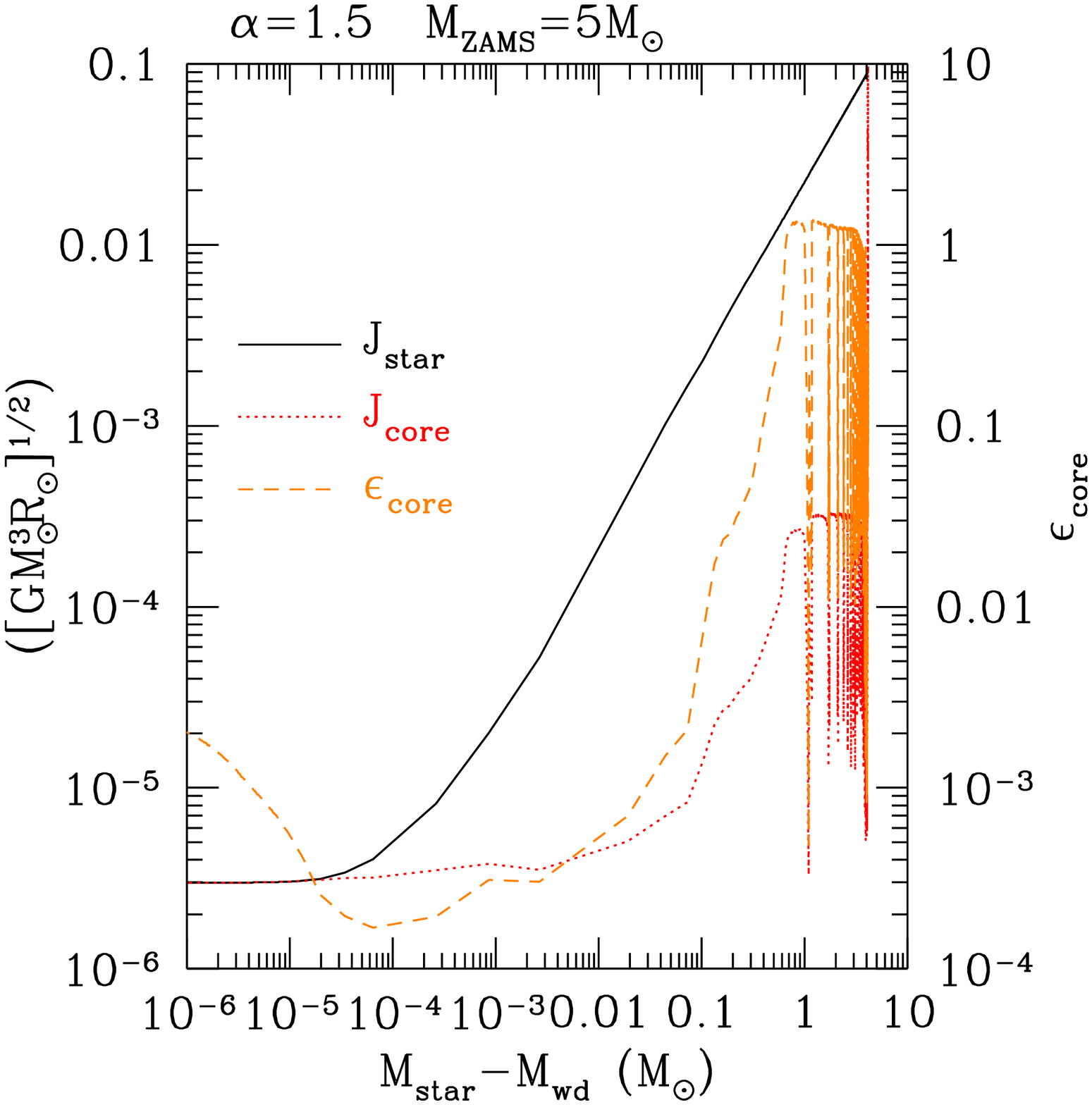}
\vskip 0in
\caption{Total spin angular momentum of AGB star during the contraction of its hydrogen envelope (black solid line)
and the angular momentum of the core (red dotted line).  Top panel:  $1\,M_\odot$ model, including both the initial
envelope contraction and the re-expansion following a late helium shell flash.  Bottome panel:  $5\,M_\odot$ model.
We show for comparison the dimensionless core moment of inertia $\epsilon_{\rm core} = I_{\rm core}/(2/3)M_{\rm benv}R_{\rm benv}^2$
(orange short-dash line).}
\vskip .2in
\label{fig:jvsm}
\end{figure}

The coefficient $\epsilon_{\rm core}$ is very small, but a key consideration is its relative
size compared with the fraction $M_{\rm env}/M_\star$ of the stellar mass that is contained in 
the envelope.  We find that $\epsilon_{\rm core}$ ranges over $\sim 3\times 10^{-5}-10^{-3}$ during
the superwind and post-AGB phases.  The envelope of the $1\,M_\odot$ model completes its first contraction 
while it retains $\sim 10^{-3}\,M_\odot$, but following its re-expansion and re-contraction this mass has dropped
to $M_{\rm env} \sim 10^{-4}\,M_\odot$.  In the $5\,M_\odot$ model, the envelope experiences a single contraction when its mass
drops below $\sim 10^{-4}\,M_\odot$.  See Figure \ref{fig:jvsm}.

Consider first the stage(s) when the envelope is still inflated, and the core radius $R_{\rm benv}$ changes only
slowly.  Then
\ba
{dJ_\star\over dt} &=& {2\over 3}R_\star^2\Omega(R_\star){dM_\star\over dt} \nn
                   &=& {2\over 3}R_{\rm benv}^2\Omega_{\rm benv} {dM_{\rm env}\over dt} + I_{\rm eff} {d\Omega_{\rm benv}\over dt}.
\ea
and
\be
{1\over\Omega_{\rm benv}}{d\Omega_{\rm benv}\over dt} = {dM_{\rm benv}/dt\over M_{\rm env} + \epsilon_{\rm core} M_{\rm benv}}.
\ee
Changes in the radius of the star do not enter into these expressions as a result of the uniform
specific angular momentum profile in the envelope.  

The core mass increases relatively slowly compared with the rapid drop in $M_{\rm env}$ during the peak of
the superwind.  So $\Omega_{\rm benv}$ increases only slightly up to the contraction of the envelope, where
it takes the value $\Omega_{\rm benv,col}$.  Beyond that point, the entire star rotates nearly as a solid body.
A wind from its surface, which carries away a mass $\Delta M_\star$ causes a net spindown
\be
{\Omega_{\rm benv,col}'\over\Omega_{\rm benv,col}} \simeq \exp\left[-{1\over\epsilon_{\rm core}}{\Delta M_\star\over
M_\star}\right].
\ee
This works out to a factor $\sim 0.2$ decrease in $\Omega_{\rm benv}$ in the $1\,M_\odot$ model at the onset of the 
late helium shell flash.

The key step in the dramatic spindown of the post-AGB star involves the re-expansion of the envelope.  This occurs
by the transfer of $\sim 10^{-3}\,M_\odot$ of hydrogen-rich material from the radiative part of the star into a rejuvenated convective layer, 
all occuring at nearly constant total angular momentum.  Now the core rotation rate (still equal to the rotation rate at the base
of the envelope due to a tight magnetic coupling) decreases to
\be\label{eq:omegaf}
{\Omega_{\rm benv}\over\Omega_{\rm benv,col}'} = {\epsilon_{\rm core}M_{\rm benv}\over M_{\rm env} + \epsilon_{\rm core}M_{\rm benv}} \sim 
{\epsilon_{\rm core}M_{\rm benv}\over M_{\rm env}} \sim {1\over 30}.
\ee
(Although the star contracts by a factor $\sim 10$ in between the initial envelope contraction and the late
helium shell flash, $\epsilon_{\rm core}$ returns to its pre-contraction value after the re-expansion of the envelope.)
The rotation rate of the core remains nearly constant at the value (\ref{eq:omegaf}), as most of the rejuvenated envelope is expelled.

The net spindown from the initial contraction of the envelope works out to $\Omega_{\rm benv}/\Omega_{\rm benv,col} \sim 1/150$.  
This behavior is neatly encapsulated by the trend of $J_\star$ and $J_{\rm core}$ with the mass $M_\star-M_{\rm wd}$ remaining to be expelled 
(Figure \ref{fig:jvsm}).  

\hfil

\subsection{Minimal Magnetic Flux Enforcing \\ Strong Core-Envelope Coupling} \label{s:coupling_flux}

Here we evaluate the minimal magnetic flux that must thread the core-envelope boundary in order for
i) the outer radiative core to remain in solid rotation; and ii) the rotation of the inner core to remain coupled 
to the outer core and inner envelope.  A related (although less quantitative) estimate has been made by \cite{sp98} for
the central cores of massive stars during the later stages of nuclear burning.

First consider the outer core.  Although it contains a small fraction of the core mass, it
can contribute significantly to the moment of inertia.  We define a characteristic magnetic flux 
threading a single hemisphere by setting $B_r/\sqrt{4\pi \rho(R_{\rm benv})} = |v_r - dR_{\rm benv}/dt|$:
\be\label{eq:phi_solid}
\Phi_{\rm solid} = \pi R_{\rm benv}^2 |v_r - dR_{\rm benv}/dt| \sqrt{4\pi\rho(R_{\rm benv})}.
\ee 

Next consider changes in the angular momentum of the entire core, which are sensitive to the evolving
rotation profile of the envelope.  The core angular momentum 
responds to the applied Maxwell stress $B_rB_\phi/4\pi$ according to
\be
I_{\rm core}{d\Omega_{\rm core}\over dt} \sim 
\int\frac{B_r B_{\phi}}{4\pi} R_{\rm benv} \sin\theta \cdot 2\pi R_{\rm benv} ^2 \sin\theta d\theta.
\ee
Taking a second time derivative, making use of the induction equation, we get
\be
I_{\rm core} \frac{d^2\Omega_{\rm core}}{dt^2} 
 \sim \frac{1}{2} \int B_r^2 \frac{\partial\Omega}{\partial r} R_{\rm benv} ^4\sin^3\theta d\theta.
\ee
Here the poloidal magnetic field is assumed to be purely radial, so that $\partial_r(r^2B_r) = 0$, and the
integral is evaluated at radius $R_{\rm benv}$.  Integrating over $\theta$ then gives
\be
I_{\rm core} \frac{d^2\Omega_{\rm core}}{dt^2} \sim \frac{2}{3} B_r^2 R_{\rm benv} ^4 \frac{\partial\Omega}{\partial r}.
\ee
We take the rotation rate to vary according to
\be
\Omega(r,t) \sim \Omega_{\rm benv}  \left( \frac{R_{\rm benv} }{r} \right)^{2} \exp\left[-\frac{t}{\tau_{\Omega}} \right].
\ee
Then to avoid spindown on a timescale $\tau_\Omega$, the magnetic flux must exceed
\ba\label{eq:phicoup}
\Phi_{\rm couple} &=& \pi R_{\rm benv}^2 B_r \sim \sqrt{\frac{3}{4}} \pi \frac{I_{\rm core}^{1/2} R_{\rm benv} ^{1/2}} 
{\tau_{\Omega}} \\ \nonumber
 &\simeq& 7.1 \times 10^{20} \ {\rm Mx} \left( \frac{I_{\rm core}}{10^{-4}M_{\odot}R_{\odot}^2} \right)^{1/2} \\\nonumber
  && \quad\quad \left( \frac{R_{\rm benv} }{R_{\odot}} \right)^{1/2} \left( \frac{\tau_{\Omega}}{10^3 {\rm yr}} \right)^{-1}.
\ea
We note that $\tau_\Omega$ is controlled by the rate at which material is ejected from the envelope during
the final superwind phase, and as a result $\Phi_{\rm couple}$ can in principle be larger or smaller than 
$\Phi_{\rm solid}$ (which is determined by the requirement that the core itself remain in solid rotation).

\begin{figure}[!]
\figurenum{16}
\epsscale{1.0}
\plotone{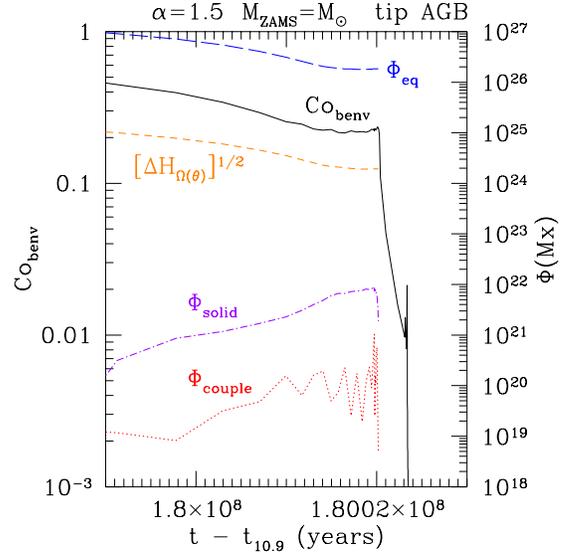}
\caption{Comparison between magnetic flux $\Phi_{\rm couple}$ (Equation (\ref{eq:phicoup}), red dotted line)
that maintains core-envelope coupling on the AGB, flux $\Phi_{\rm solid}$ (Equation (\ref{eq:phi_solid}), purple dot-dashed line)
that maintains solid rotation in the outer core against the inward mass flow,
and flux $\Delta{\cal H}^{1/2}$ generated by the $\Omega(\theta)$ dynamo process on radial drift time (orange short-dashed line).
Blue long-dashed line:  magnetic flux corresponding to a poloidal field in equipartition with the convective motions,
$\Phi_{\rm eq} = \pi R_{\rm benv}^2 \sqrt{4\pi \rho(R_{\rm benv}) v_{\rm con}^2}$.
Solid black line: Coriolis parameter at base of envelope.  We argue that the strong dependence of the dynamo-generated field
on ${\rm Co}_{\rm benv}$ allows for core-envelope decoupling below ${\rm Co}_{\rm benv} \sim 0.1-0.3$.  Fluxes are only
plotted up to the final thermal pulse: afterward $dM_{\rm benv}/dt < 0$ and the coupling of the outer core
to the envelope is less sensitive to the instantaneously generated magnetic field.}
\vskip 0in
\label{fig:Co_base_and_Phi_couple_Phi_dynamo}
\end{figure}

\begin{figure}[!]
\figurenum{17}
\epsscale{1.0}
\plotone{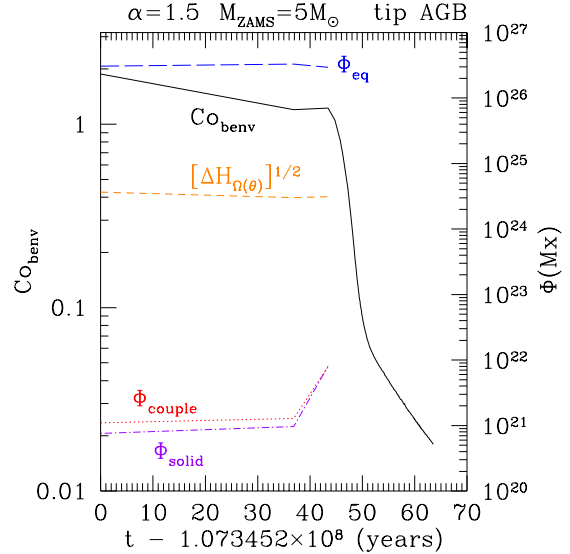}
\caption{Similar to Figure \ref{fig:Co_base_and_Phi_couple_Phi_dynamo}, but for the $M_{\rm ZAMS} = 5\,M_\odot$ model.}
\vskip .2in
\label{fig:Co_base_and_Phi_couple_Phi_dynamo_time_shift}
\end{figure}

These estimates of the magnetic flux that will enforce a tight rotational coupling between core and envelope
are compared in Figures \ref{fig:Co_base_and_Phi_couple_Phi_dynamo} and 
\ref{fig:Co_base_and_Phi_couple_Phi_dynamo_time_shift} with the radial flux $(\Delta{\cal H})^{1/2}$ 
that accumulates at the core-envelope boundary.  This
increment $\Delta{\cal H}$ is calculated using Equation (\ref{eq:helicitye}), taking the timescale
for helicity accumulation to be the minimum of the radial drift time $\ell_P/|v_r^{\rm nuc}|$ and $\tau_\Omega$.   

The larger value of the dynamo-generated flux supports our choice of ${\rm Co}_{\rm benv} < 1$ for 
core-envelope decoupling.  Recall also from Section \ref{s:bseed} that
a tiny poloidal magnetic field -- corresponding to $10^{-4} \sqrt{4\pi \rho(R_{\rm benv})} v_{\rm con}$ 
during the later stages of the AGB -- will seed a toroidal field in the tachocline 
that is strong enough to interact buoyantly with the layers above (see Figure \ref{fig:B_seed_TAGB}).

\hfil

\section{Magnetic Field Emergence and Decay}\label{s:B_field_emergence_decay}

We now consider the emergence of magnetic fields from post-AGB stars, and their subsequent decay.  The lifetime of the
visible magnetic field at the surface of the star depends on the thickness of the magnetized layer, which in turn 
depends on the mass of the progenitor and the mass loss rate during the expulsion of the hydrogen envelope.  

The strong dredge-up of helium that is experienced by stars of mass $> 2.3\,M_\odot$ leaves only
a thin mass shell to be processed on the AGB.  The remainder of the C/O material is generated during core He burning,
when the central convective core is decoupled from the exterior and therefore must conserve ${\cal H}$.  
On the other hand, the helium generated on the RGB is not dredged up in solar-mass stars, and so
magnetic fields deposited during both the red giant and asymptotic giant phases contribute directly to the remnant
WD field.

We found that mass loss on the AGB can push the convective envelope below the threshold
for dynamo activity (Figures \ref{fig:Co_base_and_J_star_J_core_TAGB_closer_solar} and
\ref{fig:Co_base_and_J_TAGB_closer_5M_solar_time_shift}).  In this case, a thin outer layer of the remnant is initially
unmagnetized;  a strong magnetic field only emerges by ohmic diffusion.  We do not consider any contribution to the magnetic
field from the contraction of the final $10^{-4}\,M_\odot$ of envelope material, which might be rapidly rotating as
the result of anisotropic mass loss \citep{Spru1998}.


The timescale for flux emergence, and the asymptotic decrease in the trapped magnetic flux,
are easily obtained from the induction equation,\footnote{Hall drift does not modify an
axisymmetric magnetic field in a fluid star, and so we consider only ohmic effects here.}  
\be \label{eq:induction}
\frac{\partial {\bf B}}{\partial t} = - \nabla \times (\eta\nabla \times {\bf B}).
\ee 
Here $\eta = c^2/4\pi \sigma$ is the magnetic diffusivity expressed in terms of electrical conductivity $\sigma$ and speed of light $c.$  
For degenerate matter in the liquid state the conductivity can be estimated using \citep{yacku1980,itohmii1983}, 
\be\label{eq:sigma}
\sigma = 8.5 \times 10^{21} {\rm s}^{-1} \ \frac{1}{\Lambda_{ei} \left<Z\right>} \frac{x^3}{1 + x^2}.
\ee
Here $\Lambda_{ei} \simeq 1$ is the Coulomb logarithm; $\langle Z\rangle = \mu_e \sum X_i Z_i^2 / A_i$, with $X_i$ the mass fraction, 
$Z_i$ and $A_i$ the nuclear charge and mass of species $i$, and $\mu_e$ is the mean molecular weight per electron. 
Finally $x = p_F / m_e c$ measures the Fermi momentum of the electrons (of mass $m_e$). 
A good approximation to the diffusion timescale through a pressure scale height $l_P$ is
\be \label{eq:t_diff_l_P}
t_{{\rm diff},l_P} \simeq \frac{4\pi \sigma l_P^2}{c^2}.
\ee

An analytic approximation to $t_{{\rm diff},l_P}$ is easily obtained in a thin outer 
shell of mass $\Delta M$ that is supported by the pressure of non-relativistic electrons,
\be
P = K\rho^{5/3} = \frac{(3\pi^2)^{2/3}}{5} \cdot \frac{\hbar^2 }{m_e ( \mu_e m_u )^{5/3}} \rho^{5/3},
\ee
where
\be
P \simeq g{\Delta M\over 4\pi R_{\rm wd}^2} = {GM_{\rm wd}\Delta M\over 4\pi R_{\rm wd}^4}
\ee
in a nearly uniform surface gravitational field $g$.  The integration of the equation of hydrostatic equilibrium gives 
a pressure scale height $l_P = (2/5)(R_{\rm wd}-r)$ and density $\rho(l_P) = (gl_P/K)^{3/2}$.  
Substituting $\sigma$ from equation (\ref{eq:sigma}) leads to
\ba \label{eq:t_diff_analytic}
t_{\rm diff,analytic} &\sim& 1.8 \times 10^{10} \ {\rm years} \left(\frac{\Delta M}{M_\odot}\right)^{7/5} 
 \left(\frac{ M_{\rm wd}}{M_\odot}\right)^{-3/5} \nn
	&& \left(\frac{R_{\rm wd}}{10^{-2}R_\odot}\right)^{-8/5}
   \left(\frac{Y_e}{0.5}\right)^{2} \left(\frac{\left<Z\right>}{6}\right)^{-1} \Lambda_{ei}^{-1}. \nn
\ea
From this expression it is easy to see that the diffusion time through a shell of fixed $\Delta M$
depends weakly on the total WD mass.

\begin{figure}[!]
\figurenum{18}
\epsscale{1.0}
\plotone{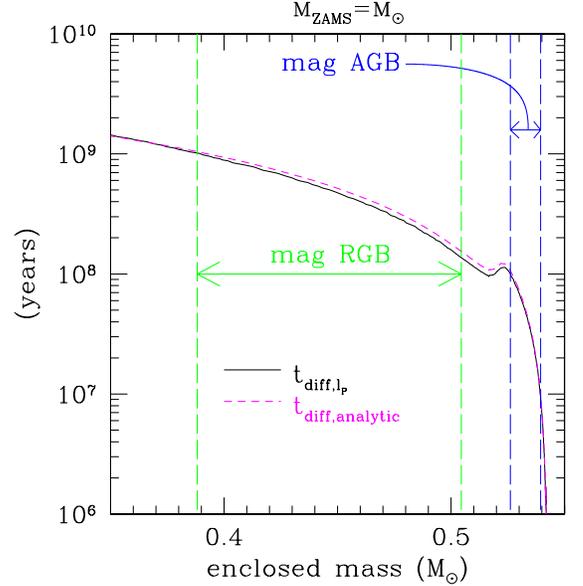}
\caption{Ohmic diffusion time in magnetized layers of $0.55\,M_\odot$ WD remnant of $1\,M_\odot$ star, 
as given by Equations (\ref{eq:t_diff_l_P}) (black line) 
and the analytic fit (\ref{eq:t_diff_analytic}) (magenta dashed line).
Mass shells passing through the dynamo-active layer on the RGB (AGB) are bounded by green (blue) dashed lines (when the star has absorbed a 
Jupiter-mass planet).
The hump in the diffusion times close to the surface is caused by a shift in composition and electrical conductivity.}
\vskip .2in
\label{fig:T_Ohm_diffusion_mass_even_earlier_model}
\end{figure}

\begin{figure}[!]
\figurenum{19}
\epsscale{1.0}
\plotone{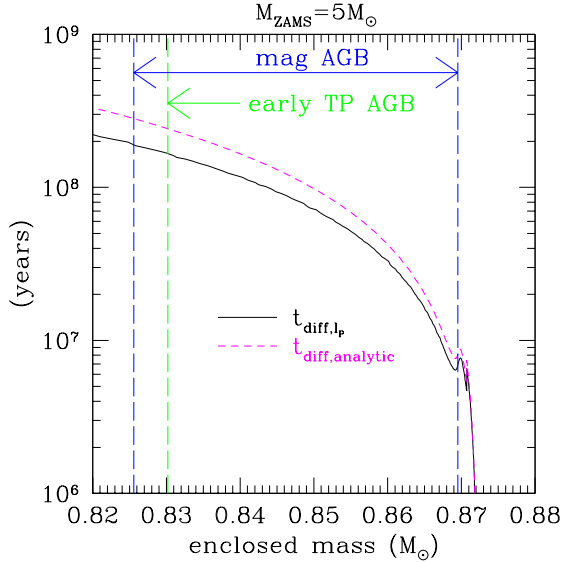}
\caption{Similar to Figure \ref{fig:T_Ohm_diffusion_mass_even_earlier_model}, but for $0.87\,M_\odot$ WD remnant of$5\,M_\odot$ star.
Mass processed during the tail of the first thermal pulse, which produces the majority of the magnetic flux (Figure 
\ref{fig:Delta_H_sqrt_TP-AGB_5M_solar}) is bounded by the dashed green line.}
\vskip .2in
\label{fig:T_Ohmic_diffusion_B_layer_mass_non_degenerate}
\end{figure}

The diffusion time through the magnetized surface layer of our 0.55 and $0.87\,M_\odot$ model WDs is shown in 
Figures \ref{fig:T_Ohm_diffusion_mass_even_earlier_model} and \ref{fig:T_Ohmic_diffusion_B_layer_mass_non_degenerate}.
The longer diffusion time in the $0.55\,M_\odot$ remnant is due to the larger magnetized mass.

A direct integration of the induction equation is straightforward if we ignore the continuing hydromagnetic adjustment 
that must accompany ohmic diffusion.  Making the same approximation of a geometrically thin magnetized layer, one has
\be
{\partial \Phi_{r,\phi}\over\partial t} = {c^2\over 4\pi\sigma(r)}{\partial^2 \Phi_{r,\phi}\over\partial r^2}
\ee
for both the hemispheric poloidal flux $\Phi_r(r) \simeq 2\pi R_{\rm wd}^2\int d(\cos\theta)  B_r(r,\theta)$,
and the toroidal flux $\Phi_\phi(r,\theta) \simeq 2\pi R_{\rm wd}\int_r^{R_{\rm wd}} dr' B_\phi(r',\theta)$ (defined
locally in $\theta$).

\begin{figure}[!]
\figurenum{20}
\epsscale{1.0}
\plotone{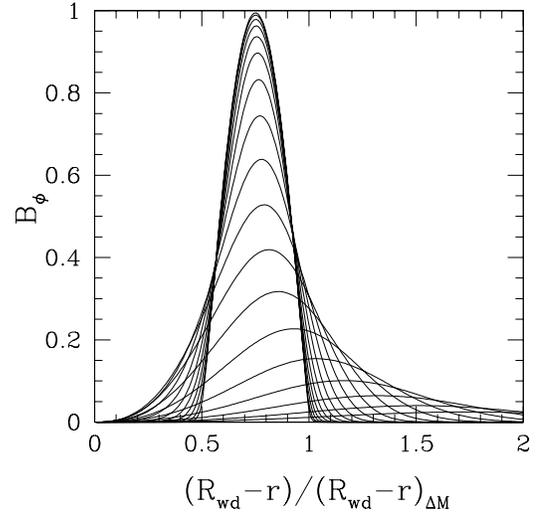}
\plotone{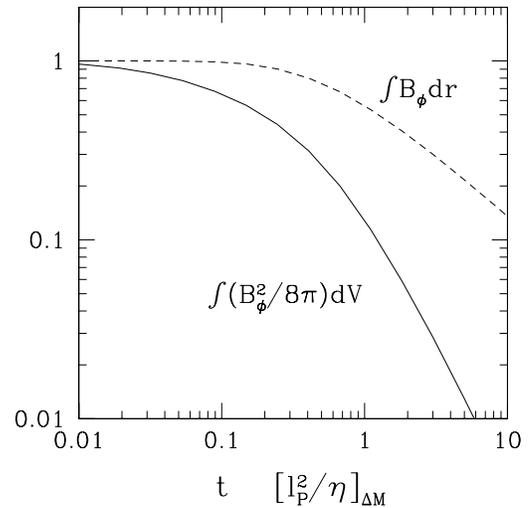}
\vskip 0in
\caption{Top panel:  ohmic diffusion of toroidal magnetic field intially concentrated at depths $0.5\Delta r$ to $\Delta r\ll R_{\rm wd}$
below the surface of a cold white dwarf.  Here $\Delta r$ is the maximum depth of the magnetized layer of mass $\Delta M$.
Bottom panel:  decrease of the toroidal (or radial) magnetic flux threading this layer, along with the
toroidal magnetic energy, both normalized to the initial values.  
This calculation does not allow for an interchange between toroidal and poloidal fluxes
resulting from a hydromagnetic instability that might accompany ohmic drift.}
\vskip .2in
\label{fig:fdecay}
\end{figure}

The decay of the flux is shown in Figure \ref{fig:fdecay}, along with the toroidal magnetic energy.  We choose a 
simple initial magnetic configuration with toroidal flux concentrated between depth $0.5\Delta r$ and $\Delta r$, 
where $\Delta r = R_{\rm wd}-r(\Delta M)$ is the depth of the magnetized layer.  Both the toroidal
and poloidal fluxes decrease by a factor $\sim 1/7$ over a time interval 10 times longer than the local decay time
$\ell_P^2/\eta(\Delta r)$, as measured at depth $\Delta r$.


At the top of the magnetized layer of both model WDs, $t_{{\rm diff},l_P}\sim 10^7$ yr.  
We conclude that the magnetic field, although initially buried, will emerge at a moderate age in both intermediate-mass and
high-mass white dwarfs.  

In the $0.55\,M_\odot$ remnant of the solar-mass star 
$t_{{\rm diff},l_P}$ extends up to $\sim 1-2 \times 10^{9}$ yr at the base of the magnetized layer; whereas it reaches a maximum 
$\sim 2\times 10^8$ yr in the $0.87\,M_\odot$ remnant of the $5\,M_\odot$ star.  
Although essentially all WD progeny of intermediate-mass stars will have strong buried toroidal magnetic fields -- due to the
high angular momentum remaining at the end of the MS -- some decay in this buried field is expected at ages exceeding $\sim 1$ Gyr.  

These ohmic timescales depend on the thickness of the C/O layer accumulated on the thermally pulsating AGB.  The WD masses
we have obtained correspond to a normalization $\eta_B = 0.05$ to the mass-loss rate on the AGB using the formula
of \cite{bloe1995}.  There are some suggestions of a lower normalization 
(e.g. $\eta_B \sim 0.01$, based on the abundance of Li-rich giants in the LMC: \citealt{ventura00}), 
which implies a  larger final C/O core.  For example, increasing the remnant mass of the $5\,M_\odot$ progenitor
to $\sim 0.90\,M_\odot$  results in an ohmic diffusion time $\sim 4\times 10^8$ yr at the base of the magnetized layer).   
One sees from  Figure \ref{fig:fdecay} that any toroidal flux buried in such a WD will have decreased by a factor
$\sim 1/5$ from its post-AGB value by an age $3\times 10^9$ yr.

Whether this decay of a buried field corresponds to significant decay of the surface dipole magnetic field -- or even
to growth -- remains uncertain because the two components do not evolve independently.  Ohmic diffusion of the magnetic field
is so slow that the magnetic field easily makes a continuing hydromagnetic adjustment to something close to 
magnetostatic equilibrium.  An interchange between the toroidal and poloidal components is possible
without any change in ${\cal H}$.



\section{Summary and Comparison with Alternative Theoretical Approaches} \label{s:conclusions}

We have investigated how the rotation and magnetism of a giant star respond
to the inward advection of a small amount of angular momentum by deep 
convective plumes in the extended envelope.  Some features of the 
dynamo process operating near the core-envelope boundary are unique to giants:
namely those driven by the intense radiation flux and the 
rapid inward drift of material to the burning shell(s).  The buoyancy that
is induced by radiative heating of magnetized material strongly enhances 
the rate of mixing between core and envelope.  

Even a weak helical magnetic field enforces a rotational coupling between
core and envelope, and maintains nearly solid rotation in the core.  
The inflow of mass into the core is accompanied by an
inflow of magnetic helicity, which is needed to sustain a stable, large-scale
magnetic field in the white dwarf remnant.  The compensating helicity is
lost through the surface of the star.

This strong core-envelope coupling is easily maintained in subgiants
and core He burning stars, as we found in Paper I.  It is also sustained
near the tips of the RGB and AGB in isolated intermediate-mass stars
if angular momentum is distributed broadly within the slowly 
rotating parts of the envelope.  We found that the threshold for dynamo
action is easily reached when $\Omega \propto r^{-2}$ at ${\rm Co} \lesssim 1$.

Injection of angular momentum from a planetary or stellar companion
is needed to maintain a magnetic core-envelope coupling during the giant 
expansion of late-type stars that lose most of their initial spin to magnetized
winds (corresponding to  $M_{\rm ZAMS} \lesssim 1.3\,M_\odot$).

Although a strong coupling does not depend on angular momentum injection
in intermediate-mass stars, the remnant magnetic field may be significantly
enhanced by the absorption of a massive planet or brown dwarf, due
to the relatively low mass of the outer shell that becomes magnetized.

These results are broadly consistent with an increased incidence of 
strong magnetism in i) massive WDs, and ii) accreting WDs 
in CVs, which experienced a tidal interaction before and during a 
common-envelope phase.  

The strength of the remnant dipole magnetic field is directly tied to the magnetic helicity
that has accumulated in the hydrogen-depleted core. 
Helicity growth depends on a combination of  i) efficient angular momentum
transport by poloidal magnetic fields through the outer core; and ii) a persistent radial
angular velocity gradient in the tachocline that is driven by angular inhomogeneities
in the rotation rate in the surrounding convective envelope.  The helicity
flux is roughly proportional to the rotational kinetic energy of the inner
envelope.

The WD dipole fields so obtained can exceed $10^7$ G in both
the $0.55\,M_\odot$ WD remnant of a solar-mass star that absorbs a Jupiter,
and the $0.87\,M_\odot$ remnant of a $5\,M_\odot$ star that has no binary interaction 
(Figure \ref{fig:B_WD_three_dynamos_vs_L_orb}).  
In the former case, the engulfment of companion much more massive than Jupiter requires
exceptional circumstances, and this field can be considered a approximate upper limit. 
In the latter case, the 
engulfment of much more massive companion, even $\sim0.1M_\odot$, may be significantly more
common.  This results in an increased remnant field, reaching as high as $\sim10^8$ G 
(Figure \ref{fig:B_WD_high_mass}).

These field strengths do not approach the strongest measured in WDs
($\sim 10^9$ G), which however are very rare.  The strongest fields
may therefore originate in more extreme merger events such
as the collision of two WDs (e.g. \citealt{gb12}).

The engulfment of an Earth-mass planet by an evolved solar-mass star, which barely augments the angular momentum of 
the star, results in a WD dipole field of $\sim10^6$ G. In this case, we find that most of the magnetic helicity is deposited
during a brief interval at the start of core He burning.  The star falls below the threshold for dynamo activity near the tips of the 
RGB and AGB.  Even a mild enhancement of wind-driven angular momentum loss, or a softening of the rotation profile in the envelope, would 
eliminate most of this MG magnetic field.

The magnetic field generated on the giant branches is initially buried in the WD (in cases where
AGB core and envelope decouple before the final thermal pulse) but diffuses 
ohmically outward over the first $\sim 10^7$ yr.  A strong internal
toroidal field begins to decay at an age $\sim 4\times 10^8$ yr
in a $\sim 0.9\,M_\odot$ WD, as compared with $\sim 2\times 10^9$ yr
in a $0.55\,M_\odot$ WD.  We find that the net decay is by a factor $\sim 0.2$ 
at an age $\sim 3\times 10^9$ yr in a $0.9\,M_\odot$ WD.  The initial
decay of the toroidal field may be accompanied by transient growth of
the external dipole, which is possible at constant internal magnetic helicity.
On the observational side, the compilation of magnetic WDs in \cite{kepletal2013} 
only extends to an age $\sim 1$ Gyr, and the incidence
of magnetism in older WDs remains uncertain.  

We find that the spin angular momentum of the remnant WD of a 
late-type star can, in some circumstances, 
{\it anti-correlate} with the angular momentum of the star during the AGB phase.  
In stars that retain only a moderate angular momentum at the end 
of the AGB (less than the orbital angular momentum of Jupiter),
this decoupling sets in before the final helium shell flash.
The final WD spin then depends most directly on the critical Coriolis parameter 
below which the envelope dynamo fades away in the inner envelope.
In the case of absorption of a Neptune-mass planet during the post-MS
evolution of a $1\,M_\odot$ star, $P_{\rm wd}$ lies between 0.1 and 1 d. 
This residual spin rate is a lower bound, as it does not include a
contribution from a final anisotropic contraction of the envelope
\citep{Spru1998}.  

The situation changes if the star is massive enough to retain most of
its natal angular momentum, or if it absorbs a Jupiter-mass planet
or brown dwarf.  Then core and envelope should remain coupled beyond
the final helium shell flash, and spindown of the core can be directly
related to a transient reduction in core mass.  The angular momentum that is retained
by the core during the contraction of the envelope depends on whether 
the core experiences a late re-expansion due to a delayed helium shell
instability.  A transfer of mass $M_{\rm env} \gg \epsilon_{\rm core}M_\star$
to the envelope, combined with ejection of most of this envelope material,
implies strong spindown of the core.  
The net result is that the final WD spin period can exceed a year in our 
$1\,M_\odot$ model.  This final spindown was found to be significantly 
smaller in the $5\,M_\odot$ model, resulting in $P_{\rm wd}$ shorter than a day,
because of the smoother transition to the post-AGB phase.

\subsection{Some Outstanding Issues}

{\it Sensitivity to strength of angular-momentum pumping in giant envelope.}  The Kepler 
asteroseismological data give a calibration of the envelope rotation profile in subgiants -- as
investigated in detail in Paper I -- but not in stars larger than $\sim 10\,R_\odot$.  Here we 
have considered the consequences of uniform $dJ/dM$ in the slowly rotating parts of the envelope.
This zone with ${\rm Co} < 1$ extends deep into the star near the tips of the RGB and AGB, unless 
the star interacts with a relatively massive binary companion.  A rotation profile $\Omega \propto r^{-2}$
in the outer envelope is supported by the numerical simulations of \cite{brunp2009}.  

The orbital angular momentum of a Jupiter can be compared with the 
minimum rotational angular momentum that will sustain a dynamo near the tip of the AGB.
Starting the planet at $a_i = 1$ AU, the excess is about a factor  
$\sim 30 ({\rm Co}_{\rm benv,crit}/0.3)^{-1}$ in our $1\,M_\odot$ model.  The 
shallowest rotation profile which could maintain a minimal Coriolis parameter ${\rm Co}_{\rm benv} \sim 0.3$
in an envelope of depth $R_{\rm benv}/R_\star \sim 0.01$ is $\Omega(r) \propto r^{-4/3}$.
In this situation, the amplitude of the helicity flux into the core, which is proportional to 
$\Omega_{\rm benv}^2$ (Equation (\ref{eq:helicitye})), would be reduced by a factor $\sim 10^{-3}$.
The corresponding reduction in $B_{\rm dipole}$, by a factor $\sim 0.03$,
would still allow for fields of order MG (see Figure \ref{fig:B_WD_three_dynamos_vs_L_orb}),
but not magnetic fields $\gtrsim 10$ MG.

A related consideration is the steepness of the angular velocity profile in the inner envelope,
where ${\rm Co} \gtrsim 1$, and its effect on the upper range that is obtained for the WD magnetic field.  
In this paper we did not simply take the profile that was determined in Paper I from asteroseismic 
models of sub-giant and helium burning stars.  During these relatively compact evolutionary phases,
an inner rotation profile $\Omega(r) \propto r^{-1}$ corresponds to ${\rm Co_{benv}} \sim 10$-$30$.  This profile is consistent with
the inward pumping of angular momentum by deep convective plumes in an adiabatic envelope with gravity $g(r) \propto r^{-1}$.
A lower inner Coriolis parameter would be maintained near the tips of the RGB and AGB if the same rotation
profile were sustained there.  

In fact, the rotation profile is expected to steepen to $\Omega(r) \sim r^{-3/2}$ in the
presence of a similar convective structure, due to the contraction of the core and the steepening of the gravity profile.
For this reason, we allow the inner index in Equation (\ref{eq:ominner}) to increase to $\alpha = 3/2$ during the
expanded phase where magnetic helicity growth is concentrated, and the remnant WD spin angular momentum is determined.
The net effect is to raise the upper envelope of the magnetic field distribution in massive WDs from
$\sim 3$ MG ($\alpha=1$) to 10-30 MG ($\alpha=3/2$).  The 
limiting magnetization and spins of lower-mass WDs hardly change over
this range of $\alpha$.

{\it What happens to any core magnetic field left over from the MS phase, when exposed to 
the turbulent motions in a slowly rotating convective envelope} (${\rm Co}_{\rm benv}~\ll~1$)?
Consider taking the magnetic flux threading a WD and spreading it across the 
core-envelope boundary of the progenitor during its AGB expansion.  Even a MG magnetic field in the
WD becomes dynamically insignificant, $B_r(R_{\rm benv}) \sim 10^{-4}(4\pi \rho v_{\rm con}^2)^{1/2}$.  
Such a weak poloidal magnetic field would rapidly diffuse over the boundary
sphere, causing fluid elements that are threaded by radial field of
opposing signs to be mixed together.  In this situation 
it is difficult to see how the material flow from the envelope into the core could maintain a large-scale poloidal 
magnetic field unless it were self-consistently maintained by a dynamo process.  A numerical experiment
studying the diffusion and reconnection of such a seed field across the spherical boundary between radiative
and convective layers is feasible and would provide interesting results.

{\it Relaxation of helical magnetic field to the `maximum-dipole' configuration.}  The surface dipole magnetic of the WD
remnant has been estimated assuming that $\Phi_r \sim \Phi_\phi \sim {\cal H}^{1/2}$.   Other magnetic configurations are 
possible, especially those with relatively stronger toroidal fields, corresponding to $\Phi_\phi \gg \Phi_r
= ({\cal H}/\Phi_\phi)^{1/2}$ (\citealt{Brai2009}).  Ohmic diffusion in WD stars with strong toroidal fields 
may cause some transient growth of the dipole field, as the magnetic field relaxes to a more isotropic configuration.

{\it Dispersion in magnetic field strength in massive WDs.}  Although a greater
fraction of massive WDs exhibit strong magnetic fields than do their $\sim 0.55-0.6\,M_\odot$ cousins, a majority do not. 
Intermediate-mass stars retain more angular momentum at the end of the MS, thereby facilitating a dynamo. 
This angular momentum can, nonetheless, be enhanced by the absorption of a massive planet, brown dwarf, or
low-mass star, or a tidal interaction with a stellar companion.   Other effects causing dispersion in the
visible surface magnetic field include exchange between toroidal and poloidal magnetic fluxes, and the
time spent by the progenitor star during the thermally pulsating AGB phase.

{\it Emergence of `frozen' magnetic field generated during the pre-MS phase (or during MS core convection)?}  The outer 
$0.3\,M_\odot$ of the $0.55\,M_\odot$ WD is processed through a deep convective envelope during the 
RGB and AGB phases.  The proportion of this material which experiences rapidly rotating convection,
and develops a strong magnetic field, depends on the initial placement of planets around the star.   For
example, if the closest Neptune or Jupiter orbits at a distance $\sim 1$ AU, then only the outer half
of the processed layer is strongly magnetized.  

The ohmic diffusion time from the base of this processed layer approaches a Hubble time, independent 
of the architecture of the planetary system.  From this we conclude that the magnetization of all but
the very oldest $\sim 0.55$-$0.6\,M_\odot$ WDs is not significantly influenced by a `frozen' core magnetic field.

The assembly history is very different for the massive WD remnants of intermediate-mass stars.  Now the bulk of the
WD mass is processed during core H and He burning, which leave unaltered the magnetic helicity
stored in the stellar core.  In the case of an isolated $\sim 0.9\,M_\odot$ WD, a core magnetic field could emerge
ohmically through the outer $\sim 0.05-0.1\,M_\odot$ that is processed on the thermally pulsating AGB, but
only with a delay of $\sim 2-5\times 10^8$ yr (depending on the normalization of mass loss during this phase).
A growth in the incidence of magnetism in older WDs could therefore point to a contribution of such a `frozen' field
to the visible surface field.  (But not exclusively so, given the competing possiblity of dipole field growth
mediated by the decay of a toroidal field that is buried in a thin outer mass shell.)


{\it Implications of Binary Magnetic WDs.}  The incidence of strong magnetism (exceeding $10^7$ G
and implying classification as a `polar') exceeds $\sim 30\%$ in
short-period CVs with ages exceeding $\sim 1$ Gyr \citep{Gans2005}.   Our calculations
suggest that these fields might decay significantly in more massive isolated WDs.  Long-lived 
magnetism in old binary WDs with periods $< 2$ hr (below the `period gap') could be maintained
by the accretion of $\gtrsim 0.1\,M_\odot$ from the companion, which would have the effect of pushing
down the magnetized layers to a greater depth and a longer ohmic timescale.

\begin{figure}[!]       
\figurenum{21}
\epsscale{1.0}
\plotone{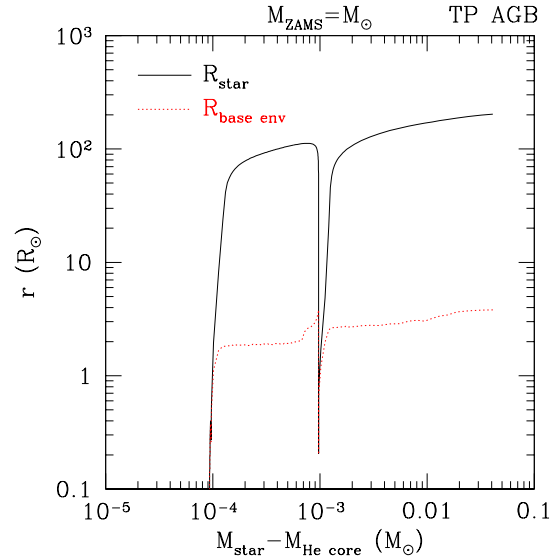}
\vskip 0in 
\caption{Dependence of AGB stellar radius $R_\star$ (black line) and base of convective envelope $R_{\rm benv}$
(dotted red line) on residual H mass.  Endpoint of $M_{\rm ZAMS} = 1\,M_\odot$ model.}  
\vskip .2in
\label{fig:AGBendpoint}
\end{figure}

\subsection{Competing effects of a collapsing low-mass ($\sim 10^{-4}\,M_\odot$) AGB envelope}  

The final contraction of an AGB envelope, beginning at a photospheric radius $\sim 10^2\,R_\odot$,
leaves behind a hydrogen layer of mass $M_{\rm env,f} \sim 10^{-4}\,M_\odot$.
Mass loss slows down significantly as the star contracts inside a radius $R_{\rm pAGB} \sim R_\odot$ (Figure \ref{fig:AGBendpoint}).

It is possible that the process driving mass loss 
is so anisotropic that the collapsing envelope carries with it a significant angular momentum,
equal and opposite to that lost during the final stages of the super-wind  \citep{Spru1998}.   
An upper bound to the angular momentum thus added to the post-AGB star can be obtained as follows.
The collapsed envelope will experience a bar instability and 
spread outward into a disk if its angular velocity exceeds 
$\sim 0.3(GM_{\rm wd}/R_{\rm pAGB}^3)^{1/2}$ at radius $R_{\rm pAGB}$.  Any angular momentum in excess of 
$J_{\rm max} \sim 0.3 M_{\rm env,f} (GM_{\rm wd}R_{\rm pAGB})^{1/2}$ is transported outward by viscous torques in the disk, and is
presumably lost in a wind.  

Setting the envelope angular momentum to this limiting value, and assuming that it is shared with the rest of the star
at radius $R_{\rm pAGB}$, we obtain a minimum spin period for the cold WD remnant, 
\be
P_{\rm wd} \geq {2\pi I_{\rm wd}\over J_{\rm max}} = 0.4\,\left({M_{\rm env,f}\over 10^{-4}\,M_\odot}\right)^{-1}
            \left({R_{\rm pAGB}\over R_\odot}\right)^{-1/2} \quad {\rm hr}.
\ee
A comparison with Figure \ref{fig:P_rot_WD_Co_crit_w_5M_o} shows that somewhat faster WD 
spins can be imparted by rotating envelope contraction than would result from a cutoff of the envelope dynamo 
before the final intense phase of mass loss.  However, the fastest WD spins may depend on a merger between two WDs.

\subsection{Impulsive Dynamo Amplification in a Merger?}

The formation of magnetic WDs in stellar binaries is considered by \cite{ToutWLFP2008}.  It is noted that the injection
of angular momentum from a binary companion (either a low-mass MS star or a substellar companion) into a giant star
will trigger strong differential rotation in its core and envelope.
This differential rotation may spark the impulsive growth of the magnetic
field in the core, further details of which have been examined by \cite{WickTF2014}.

\cite{NordWSMB2011} consider the tidal shredding of a planetary companion close to the giant core.   
They argue that this material initially forms a disk in nearly Keplerian rotation about the core, in which a magnetic field is generated by 
the magnetorotational instability.  The most promising application of this process is to isolated magnetic WDs, since it may be difficult
for a planet and a stellar companion (later the mass donor in a CV system) to co-exist on similar orbits.

The contribution made by such an impulsive dynamo to the net magnetic helicity stored in the stellar core can be compared with
the longer term effect examined in this paper.  When the companion is tidally disrupted (either within the convective
envelope, or within the rarefied mantle surrounding the burning shells), the mass $M_2$ deposited would 
be re-distributed back through the convective envelope on the thermal timescale 
\be\label{eq:tth}
t_{\rm th} \sim {GM_{\rm core} M_2 \over R_{\rm tide} L_{\rm AGB}} \sim 600 \left({M_2\over 0.1\,M_\odot}\right)
\left({R_{\rm tide}\over R_\odot}\right) \quad {\rm yr}.
\ee

As envisaged by \cite{NordWSMB2011}, a centrifugally supported disk may spread on a shorter timescale
than (\ref{eq:tth}), in which case the internal heat generated in the disk is advected around without being
radiated back into the surrounding stellar envelope.
Studies of advective accretion flows in other contexts (e.g. \citealt{bb99}) suggest that a fraction of the disk
material would be accreted deeper into the gravitational potential of the degenerate core,
which may be enough to transiently increase its rotation.

Since the energy released by H burning easily suffices to unbind the ashes from the degenerate
core, any such rapidly accreted material must remain H rich.  On the longer thermal timescale,
this nuclear energy can be radiated away, accompanied by a relaxation of the outer core to its equilibrium structure.
The mass $\delta M_{\rm nuc}$ processed by H and He burning during that relaxation is small compared with 
the total deposited mass $M_2$.  Given that a mass $M_{\rm acc}$ is accreted close to the burning radius 
$R_{\rm burn} \sim 0.1\,R_\odot$, one has
\ba
\delta M_{\rm nuc} &\sim& {GM_{\rm core}\over \varepsilon_{\rm nuc} R_{\rm burn}} M_{\rm acc} \nn
 &\sim& 2\times 10^{-3}\left({R_{\rm burn}\over 0.1~R_\odot}\right)^{-1} \left({M_{\rm acc}\over M_2}\right) M_2.
\ea
The normalization is about $\sim 10$ times larger if H burning is extinguished, 
so that only the He-burning shell remains active.  

We see that a relatively small mass is incorporated into the
C/O core before the star adjusts to a new equilibrium structure.  Given that the magnetic coupling between core and envelope
is sustained, most of the added angular momentum in this new equilibrium state will be stored in the outer envelope.

The large-scale Maxwell stress $B_r B_\phi/4\pi$ in the outer core may easily increase by a factor $\sim 10^4$ in the
immediate aftermath of the tidal disruption, when the rotation frequency increases by $\gtrsim 10^2$.  However, as the
envelope dredges up the injected angular momentum, we expect the Maxwell stress to adiabatically adjust downward, since
little mass flows into the core.  Such a downward adjustment could be avoided if the magnetic field lines
became decoupled from the envelope, but that does not seen consistent with the rapid buoyant motion of such a
strong magnetic field (see Equation (\ref{eq:buoy})).  

From this perspective, the {\it impulsive} growth of the magnetic field following the tidal disruption of a low-mass star or 
substellar companion may leave little permanent mark on the magnetization of the C/O core.   A merger during the early 
post-MS expansion could induce some direct hydrodynamic mixing with the hydrogen-depleted core, but only when the accumulated
helium mass was still well below the final WD mass.   

We argue that the dominant effect of such a merger is, instead, to i) push the envelope above the threshold for a gentler and more persistent 
dynamo process in low-mass giant stars, $M_{\rm ZAMS} \lesssim 1.3\,M_\odot$; and ii) to augment the inward flux of magnetic helicity in stars
which remain rapidly rotating at the end of the MS.  

The merger of two WDs (or of a WD with the core of an evolved companion) provides the main counter-example 
to these conclusions.   Now hydrogen-depleted material in the more massive WD experiences direct
hydrodynamic mixing with the tidally shredded remnants of the companion.  The rate of such events appears to
be relatively small, and may therefore only accomodate the most strongly magnetized WDs.

\acknowledgements  We would like to thank David Arnett and Peter Goldreich for conversations.  This work was supported by NSERC.

\begin{appendix}

\section{Twisting of a Radial Magnetic Field} \label{s:twist}

Here we construct expressions describing the large-scale flow of magnetic 
helicity in the evolving core of a giant star.   The rotation profile
and magnetic field are assumed to be axisymmetric, but departures from
reflection symmetry about the rotation axis are considered.  

Our treatment is simplified 
by considering a weak radial magnetic field with angular flux profile
\be
\Phi_r(\theta) = \int_0^\theta \sin\theta' r^2B_r(\theta') d\theta'
\ee
and corresponding potential
\be
A_\phi(\theta,r) = {1\over r\sin\theta}\Phi_r(\theta).
\ee
In what follows, we therefore neglect any radial change in $\Phi_r(\theta)$.  

A dynamo operating near the convective boundary will generally supply a 
poloidal field of a fluctuating sign, but the contribution to 
${\bf A}\cdot{\bf B}$ in one hemisphere is insensitive to this sign. The 
contributions from the two hemispheres generally have opposing signs.  

Consider the case where the rotation profile is reflection-symmetric,
so that convective motions enforce a certain $\partial\Omega/\partial r$ 
at polar angles $\theta$ and $\pi-\theta$.  The evolution of the
twist in the core depends on the connectivity of the poloidal magnetic field.
If both angles $\theta$ and $\pi-\theta$ are connected by the same 
(axially-symmetric) magnetic flux surface, then there is no evolution
of the net twist along this flux surface.  On the other hand, if the distribution
of magnetic flux is different across the two hemispheres, then a
given flux surface will experience a net differential winding.

The rate of change of ${\cal H}$ inside a spherical boundary of a fixed radius $r$ is
\be \label{eq:helicity}
{\partial{\cal H}\over \partial t} = -\int dS\,\hat r\cdot\left[ c\phi {\bf B} + ({\bf A}\cdot{\bf B})
{\bf v} - ({\bf A}\cdot {\bf v}){\bf B}\right],
\ee
where the bounding sphere is placed just inside the convective boundary.  

The electrostatic potential $\phi$ appearing on the right-hand side of (\ref{eq:helicity}) is driven by the mean rotation of the star. 
The corresponding electric field ${\bf E} = -(\bOmega\times {\bf r})\times{\bf B}/c$ is sourced by
\be
\phi(\theta,r) = {1\over c}\Omega(r) \Phi_r(\theta).
\ee
Differential rotation generates $A_r$ (and thence helicity) via
\be
{\partial A_r\over \partial t} = -{\partial\Omega\over\partial r}\Phi_r(\theta).
\ee
The electrostatic potential cancels off the term $A_\phi \hat\phi\cdot(\bOmega\times{\bf r})$ in Equation (\ref{eq:helicity}), and we are 
left with
\be \label{eq:helicityb}
{d{\cal H}\over dt} \;=\; {\partial{\cal H}\over \partial t} + {dR_{\rm benv} \over dt}\int dS A_\phi B_\phi
\;=\;  -\delta v_r\int dS\,A_\phi B_\phi.
\ee
Here we have subtracted off the velocity of the convective boundary from the radial flow speed $v_r$ of spherical mass shells:
\be
\delta {\bf v} = {\bf v} - {dR_{\rm benv} \over dt}\hat r.
\ee
On the AGB, $v_r$ is negative and larger in magnitude than $dR_{\rm benv} /dt$, but we include this correction for completeness. 

Because the poloidal flux surfaces are, in this treatment, fixed in position, it is useful to re-express the toroidal field in terms of a 
twist angle $\phi_B$:
\be \label{eq:bphi}
B_\phi(r,\theta) = {\partial\phi_B\over\partial r} r\sin\theta B_r.
\ee
The toroidal component of the induction equation is
\be
{\partial B_\phi\over \partial t} = B_r r\sin\theta {\partial\Omega\over\partial r} - 
{1\over r}{\partial\over\partial r} \left(r B_\phi v_r\right),
\ee
which simplifies to
\be \label{eq:phiB}
{d\over dt}\left({\partial\phi_B\over \partial r}\right) = {\partial^2\phi_B\over\partial t\partial r} + v_r {\partial^2\phi_B\over
\partial r^2} = {\partial\Omega\over \partial r} - {\partial v_r\over\partial r}{\partial\phi_B\over\partial r}.
\ee
in the case (considered here) that $\Phi_r$ only depends on latitude.

Substituting expression (\ref{eq:bphi}) into (\ref{eq:helicityb}) gives
\be \label{eq:helicityc}
{d{\cal H}\over dt} = -2\pi\delta v_r\int d\theta \Phi_r {\partial \Phi_r\over\partial\theta} {\partial\phi_B\over \partial r},
\ee
which mirrors the standard expression for a static plasma.  

In general the radial magnetic flux profile differs in shape, as well as sign,
between the two hemispheres.  A complete 
cancellation in the integral over $\theta$ in Equation (\ref{eq:helicityc}) remains possible if $\phi_B$ is independent of $\theta$.  
That would be the case if fluid stresses were to smooth out latitudinal differential rotation within each shell, so that 
$\partial\Omega/\partial r$, the source for $\phi_B$, were only a function of radius.  We see that a net flux of magnetic helicity into the 
core depends on {\it latitudinal} differential rotation -- differential rotation across flux surfaces -- which is a generic consequence of 
convection.

\end{appendix}

\end{document}